\documentclass[structabstract]{aa}  
\usepackage{graphicx}
\usepackage{lscape}
\usepackage{longtable}
\usepackage{natbib}
\usepackage[latin1]{inputenc}
\bibpunct{(}{)}{;}{a}{}{,}
\usepackage{txfonts}
\begin{document}

\title{A $\lambda=1.3$\,mm and 2\,mm molecular line survey towards M\,82 }

  \author{R. Aladro
         \inst{1,2},
         S. Mart\'in\inst{3}, J. Mart\'in-Pintado\inst{4},  R. Mauersberger\inst{5}, C. Henkel\inst{6,7}, B. Oca\~na Flaquer\inst{1,8}, M.A. Amo-Baladr\'on\inst{4}}

  \institute{Instituto de Radioastronom\'ia Milim\'etrica (IRAM),
             Avda. Divina Pastora, 7, Local 20, E-18012 Granada, Spain\\
             \email{r.aladro@ucl.ac.uk}
        \and
	Department of Physics and Astronomy, University College London, Gower Street, London WC1E 6BT, UK
	\and
        European Southern Observatory, Avda. Alonso de C\'ordova 3107, Vitacura, Casilla 19001, Santiago 19, Chile
	\and
            Centro de Astrobiolog\'ia (CSIC-INTA), Ctra. de Torrej\'on Ajalvir km 4,
E-28850 Torrej\'on de Ardoz, Madrid, Spain
       \and
        Joint ALMA Observatory, Avda. Alonso de C\'ordova 3107, Vitacura, Santiago, Chile
	\and
	Max-Planck-Institut f\"ur Radioastronomie, Auf dem H\"ugel 69, 53121 Bonn, Germany
	\and
	Astronomy Department, King Abdulaziz Univeristy, P.O. Box 80203, Jeddah 21589, Saudi Arabia
        \and
        National Centre for Radio Astrophysics, TIFR, Ganeshkhind, Pune 411007, India
            }

  \date{Received / Accepted}


 \abstract
{}
{We study the chemical complexity towards the central parts of the starburst galaxy M\,82, and investigate the role of certain molecules as tracers of the physical processes in the galaxy circumnuclear region.}
{We carried out a spectral line survey with the IRAM-30m telescope towards the northeastern molecular lobe of M\,82. It covers the frequency range between 129.8\,GHz and 175.0\,GHz in the 2\,mm atmospheric window, and between 241.0\,GHz and 260.0\,GHz in the 1.3\,mm atmospheric window.}
{Sixty-nine spectral features corresponding to 18 different molecular species are identified. In addition, three hydrogen recombination lines are detected. The species NO, H$_2$S, H$_2$CS, NH$_2$CN, and CH$_3$CN are detected for the first time in this galaxy. Assuming local thermodynamic equilibrium, we determine the column densities of all the detected molecules. We also calculated upper limits to the column densities of fourteen other important, but undetected, molecules, such as SiO, HNCO, or OCS. We compare the chemical composition of the two starburst galaxies  M82 and NGC\,253. This comparison enables us to establish the chemical differences between the products of the strong photon-dominated regions (PDRs) driving the heating in M\,82, and the large-scale shocks that influence the properties of the molecular clouds in the nucleus of NGC\,253.}
{Overall, both sources have different chemical compositions. Some key molecules highlight the different physical processes dominating both central regions. Examples include CH$_3$CCH, c-C$_3$H$_2$, or CO$^+$, the abundances of which are clearly higher in M\,82 than in NGC\,253, pointing at photodissociating regions. On the other hand, species such as CH$_2$NH, NS, SiO, and HOCO$^+$ have abundances of up to one order of magnitude higher in NGC\,253 than in M\,82.}

  \keywords{ ISM: molecules --
	galaxies: abundances
	galaxies: starburst --
        galaxies: individual: M\,82, NGC\,253--
       galaxies: nuclei --
       Radio lines: ISM}
\authorrunning{Aladro et al. (2011)}
\titlerunning{A $\lambda=1.3$\,mm and 2\,mm molecular line survey towards M\,82}
  \maketitle{}


\section{Introduction}
\label{sect.Intro}

Despite its small size, M\,82 is the brightest IRAS galaxy ($S_{100}=1351.02$\,Jy; $L_{\rm{IR}}=3\times10^{10}\rm{L_\odot}$; \citealt{Rice88,Telesco88}). Its central starburst region, of $\sim 500$\,pc size \citep{OConnell78}, has a bolometric luminosity 100 times higher than a region of equivalent size in our Galaxy  \citep{Rieke80,Lester90,OConnell95}. Almost all its luminosity arises from the central kiloparsec \citep{Lo87}, which seems to be filled with a large number of compact star clusters where high-mass stars dominate the luminosity \citep{Rieke80,Knapp80,Fuente08,Westmoquette09}. The galaxy M\,82 is considered a prototype starburst galaxy because of its high star-forming rate ($\sim9\,\rm{M_\odot}\,\rm{yr}^{-1}$, \citealt{Strickland04}), which was probably caused by a recent interaction with its companion M\,81, at a projected angular distance of 36$'$  \citep{Cottrell77,Brouillet91,Sofue98,Sun05}. The M\,82 burst of star formation is thought to be fuelled by large amounts of molecular gas ($M_{\rm{H_2}}\sim 2\times10^8\,\rm{M_{\odot}}$, \citealt{Naylor10}) still present in the central 1\,kpc region. This region shows two emission peaks opposite the dynamical centre, the north-east (NE) and south-west (SW) molecular lobes \citep{Mao00,Burillo02,Martin06a}, which are likely part of a nuclear ring of warm gas and dust heated by stars \citep{Nakai87,Dietz89}.

Apart from its enhanced activity, the proximity of M\,82 ($D$ = 3.6\,Mpc, \citealt{Freedman94}) ensures that it is one of the extragalactic sources studied in most detail at all wavelengths, from radio to X-rays. M\,82 shows a bipolar asymmetric outflow of cool (atomic and molecular), warm and hot gas \citep{Lynds63,Bregman95,Shopbell98}, which extends above and below the galactic plane to large scales, even reaching the halo \citep{Seaquist91,McKeith95,Seaquist01,Veilleux09}, and creating chimneys, supershells, and bubbles of atomic and molecular material \citep{Burillo01,Wills02}. These superwinds are generated by the nuclear starburst, which, accompanied by a high supernova rate, creates large photon-dominated regions (PDRs) in the central parts of the galaxy. At the same time, this galaxy is a strong emitter of X-rays (e.g. \citealt{Read02}) and cosmic-rays (e.g. \citealt{Veritas09}), although they do not seem to dominate the chemistry of the M\,82 nucleus as much as the UV field \citep{Fuente08}. 

The UV field in the nuclear region of M\,82 characterizes the chemistry of its molecular gas. Many species show lower abundances in M\,82 than in other nearby starburst galaxies (such as NGC\,253, NGC\,4945, or NGC\,6946). In particular, the more complex ones (with $>3$ atoms) such as CH$_3$OH, NH$_3$, or HNCO, seem to be underabundant, since they are easily dissociated by the strong UV fields \citep{Rieke80,Rieke88,Nguyen89,Takano02,Mauersberger03,Martin09a}. In particular, the HNCO abundance shows large variations of nearly two orders of magnitude among starburst galaxies \citep{Martin09a}. This suggests that the observed HNCO abundances are related to the evolutionary stage of their nuclear starbursts.

The chemistry of the M\,82 nucleus seems to be the result of an old starburst mainly dominated by the photodissociating radiation from the already-formed massive stars \citep{Fuente08}. Thus, most of the dense molecular material for future star-forming events has already been consumed in this galaxy \citep{Martin06a}. Other galaxies, notably the ones cited above, are interpreted as being at an earlier stage of a starburst, where PDRs, are significant \citep{Martin09b}, but do not yet play a major role. They are instead dominated by large-scale shocks among molecular cloud complexes, caused by cloud-cloud collisions and mass loss from massive stars \citep{Burillo00,Burillo02,Wang04,Aladro10}.

To confirm that the evolutionary states of starburst galaxies can be characterized by their chemical compositions, we performed 2\,mm and 1.3\,mm  spectral scans towards the NE molecular lobe of M\,82. In Sect.~\ref{sect.Obs}, we present the observations and the data reduction. The data analysis and the estimated uncertainties are explained in Sect.~\ref{Sect.3}. In Sect.~\ref{Section4}, we present the detected molecules as well as some important undetected species. Section~\ref{sect.comp} shows a detailed comparison of the chemistry that we found in M\,82 and NGC\,253, using a survey of similar frequency performed towards the second galaxy by \citet{Martin06b}. Finally, in Sects.~\ref{conclusions} and ~\ref{summary} we draw our conclusions and summary.

\section{Observations and data reduction}
\label{sect.Obs}

\begin{figure*}
\begin{center}
	\includegraphics[angle=0,width=\textwidth]{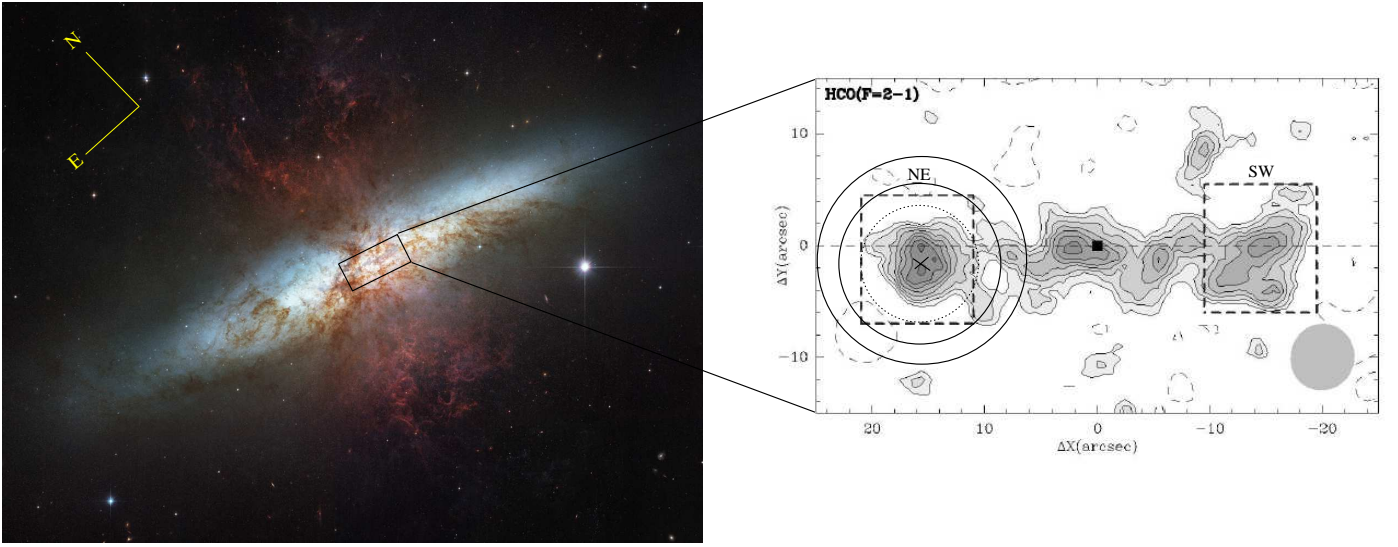}
	\caption{Left: Hubble Space Telescope (HST) image of the M\,82 galaxy taken from http://hubblesite.org (Credit: NASA, ESA, and The Hubble Heritage Team (STScI/AURA)). The right panel shows the rotated HCO\,$(2-1)$ interferometric map by \citet{Burillo02} towards the central region of the galaxy. The dynamical centre is marked with a small black square. The north-east (NE) and south-west (SW) molecular lobes are shown with two dashed squares either side of the centre. The nominal position of our observations, at an offset $(+13,+7.5)$, is marked with a cross. The beam sizes of the 2\,mm survey for the lowest and highest frequencies are marked with two solid circles. The average beam size at 1.3\,mm frequencies is marked with a dotted circle.}
\label{FigM82}
\end{center}
\end{figure*}

Observations were made with the IRAM 30-m telescope\footnote{IRAM is supported by INSU/CNRS (France), MPG (Germany) and IGN (Spain).}, located at Pico Veleta, Spain, during several periods between 2006 and 2009. The frequency survey was carried out towards the NE molecular lobe of the starburst galaxy M\,82, which has an offset position of $(\Delta \alpha,\Delta \delta)=(+13.0'', +7.5'')$ with respect to the dynamical centre ($\alpha_{2000}$\,=\,9$^h$:55$^m$:51$^s$.90, $\delta_{2000}$\,=\,$+$\,69$^o$:40$^{'}$:47$^{''}$.10; \citealt{Joy87}) (Fig.~\ref{FigM82}). At this molecular cloud complex, the dissociating radiation is claimed to be stronger than in the SW lobe and the centre regions \citep{Burillo02}. We covered the frequency range from 129.8\,GHz to 175.0\,GHz at 2\,mm wavelengths, and from 241.0\,GHz to 260.0\,GHz at 1.3\,mm. We simultaneously used the (now decommissioned) C and D receivers in single sideband mode, which provided an instantaneous bandwidth of $2\times1$\,GHz at both 2 and 1.3\,mm. As a backend, we used the 256\,$\times$\,4\,MHz filterbanks. The observations were made by wobbling the secondary mirror with a switching frequency of 0.5\,Hz and a beam throw of 220$''$ in azimuth. The data were calibrated using the standard dual load method. The rejection of the image sideband was measured for each frequency tuned, and then varied from 9\,dB to 24\,dB. The pointing was checked every one or two hours by observing planets and nearby bright continuum sources. We estimate the pointing accuracy to be $\sim3''$. The focus was always checked at the beginning of each run and during sunsets. The beam sizes ranged from 19$''$ to 14$''$ in the 2\,mm band and from 10$''$ to 9$''$ in the 1.3\,mm band. The spatial resolution ranged between 158\,kpc and 333\,kpc.

The observed spectra were converted from antenna temperatures ($T_{\rm A}^*$) to main beam temperatures ($T_{\rm MB}$) using the relation $T_{\rm MB}=(F_{\rm eff}/B_{\rm eff})\,T_{\rm A}^*$, where $F_{\rm eff}$ is the forward efficiency of the telescope (which was 0.93 and 0.88 for the 2\,mm  and 1.3\,mm bands, respectively) and $B_{\rm eff}$ is the main beam efficiency, ranging from 0.72 to 0.64 for the 2\,mm band, and from 0.50 to 0.43 for the 1.3\,mm band. Figure~\ref{FigCachos1} shows the reduced data. Baselines of order 0 or 1 were subtracted in most cases. We achieved an rms of $\sim(2-5)\,\rm mK$ across most of the 2\,mm band, and of $\sim(3-8)\,\rm mK$ for the 1.3\,mm band (for the above mentioned 4\,MHz channels corresponding to $\sim(7-9)$\,km\,s$^{-1}$ and $\sim(4-5)$\,km\,s$^{-1}$, respectively). 

\section{Data analysis and uncertainties estimates}
\label{Sect.3}

Assuming local thermodynamic equilibrium (LTE) and optically thin emission for all the detected species, we plotted Boltzmann diagrams in order to estimate the rotational temperatures ($T\rm_{rot}$) and the molecular column densities ($N_{\rm mol}$). The details of the method are explained in \citet{Goldsmith99}. The resulting rotational diagrams are shown in Fig.~\ref{FigDR} and the derived physical parameters are listed in Table~\ref{TableNT}. 

For the LTE analysis, the data were also corrected for beam dilution by $T_{\rm B}=[(\theta^2_{\rm s}\,+\,\theta^2_{\,\rm b})\,/\,\theta^2_{\,\rm s}]\,T_{\rm MB}$, where $T_{\rm B}$ is the source-averaged brightness temperature, $\theta_{\,\rm s}$ is the source size, and $\theta_{\,\rm b}$ is the beam size in arc-seconds.

\subsection{Opacity of the detected molecules}
\label{depth}
The rotational diagram method assumes optically thin lines. To conclusively prove that extragalactic lines are actually optically thin is, however, not an easy task because the ratios of hyperfine structure lines can be used only in those exceptional cases where the separation between these lines is larger than the Doppler width. This is the case for CN (e.g. \citealt{Henkel98}) and also for NO, but not for most other molecules with hyperfine structure. A second possibility is to compare line intensity ratios with known isotopic ratios: if there is no isotopic fractionation because of chemical effects (which are basically expected to occur at lower temperatures) or selected photodissociation, a line intensity ratio of the same transition of two isotopologues that is close to the isotope ratio can be taken as an indication that both lines are optically thin. The problem for extragalactic studies is: a) that isotopic lines (in particular for molecules that contain two substitutions of rare isotopes) may be too weak to be detected; b) the isotopic ratios are not a priori known and may differ from those expected in our own Galaxy (e.g. \citealt{Martin10}); and c) the sources are not resolved, hence the opacity cannot be determined from the brightness temperature and the excitation temperature. The lines of $^{12}$C$^{16}$O are the only millimeter lines for which we can be sure that these are optically thick in many extragalactic sources. The fact that all the other mm-lines in extragalactic sources tend to be more than a factor of ten weaker than CO may be an indication, but no proof, of optical thinness, since the low intensity may be attributed to a lower excitation temperature (which is a reasonable assumption because CO has a much lower dipole moment than most other molecules with detectable lines), and/or a lower beam filling factor (again something to be expected because the higher dipole species trace a denser gas component).

Nevertheless, from these arguments it is plausible to assume that all the lines observed in our survey are optically thin or at most only moderately optically thick. The strongest lines in our survey are those of CS, C$_2$H, and CH$_3$CCH. The C$^{32}$S/C$^{34}$S intensity ratio of ~13 (calculated using the $(3-2)$ transitions detected in our survey), is smaller than the isotopic ratio of 23 derived for our Galaxy \citep{Wilson94}, but there are reasons to believe that $^{34}$S may have a higher abundance in starburst galaxies than in our Milky Way \citep{Martin05}. For C$_2$H, a very stringent lower limit to its $^{13}$C substitutions was determined by \citet{Martin10}, which suggests that the lines of CCH are also optically thin in M\,82. In addition, for the case of CH$_3$CCH it is plausible that its lines are optically thin because this is a heavy molecule with a K-ladder structure where, hence, the population is spread over many energy levels.

\subsection{Source size}
On the basis of M\,82 interferometric observations  of $^{12}$CO and C$^{18}$O \citep{Weiss01}, HCO \citep{Burillo02}, HOC$^+$, and H$^{13}$CO$^+$ \citep{Fuente08}, we assumed an average source size of 12$''$ for all the molecular species detected in this survey. However, this is only a rough approximation because other species might exist in more clumpy or extended regions. To quantify the uncertainties in different source sizes for each molecule, we assumed a $\theta_{\rm s}$ variation of 50\%. We then recalculated all the column densities and rotational temperatures for $\theta_{\rm s}$= 6$''$, 12$''$, and 18$''$. The emission size is very unlikely to be either smaller or larger than these values (although interferometric maps would be necessary to prove this). We found that the uncertainty in the source size has no strong influence on the derived $N_{\rm mol}$ and $T_{\rm rot}$ parameters, limiting variations to maximal factors of 3.5 and 1.3, respectively.

\subsection{Different volumes}
Another uncertainty related to $N_{\rm mol}$ and $T_{\rm rot}$ was introduced by our assumption that all the transitions of a certain molecule arise from the same gas volume. Low-$J$ transitions might arise from more extended regions than high-$J$ transitions. In a similar way to \citet{Bayet09b}, we selected CH$_3$CCH (methyl acetylene) as a good example of a species for which we detected several transitions, because its lines are well spread over both the 2\,mm and 1.3\,mm bands. Thus, to quantify the contribution of this uncertainty, we plotted again the Boltzmann diagrams by successively decreasing the source size by 10$\%$ for each higher transition, i.e., CH$_3$CCH\,$(8-7)$ (which is the lowest transition we detected) had a $\theta_{\rm s}=12''$, while CH$_3$CCH\,$(16-15)$ (the highest transition we have) had a $\theta_{\rm s}=7.2''$. When comparing the new Boltzmann diagram results with the ones for which we applied the same source size to all the lines, we found a difference of only $\sim14\%$ in both  $N_{\rm mol}$ and  $T_{\rm rot}$.

\begin{figure*}
\centering
\includegraphics[angle=0,width=\textwidth]{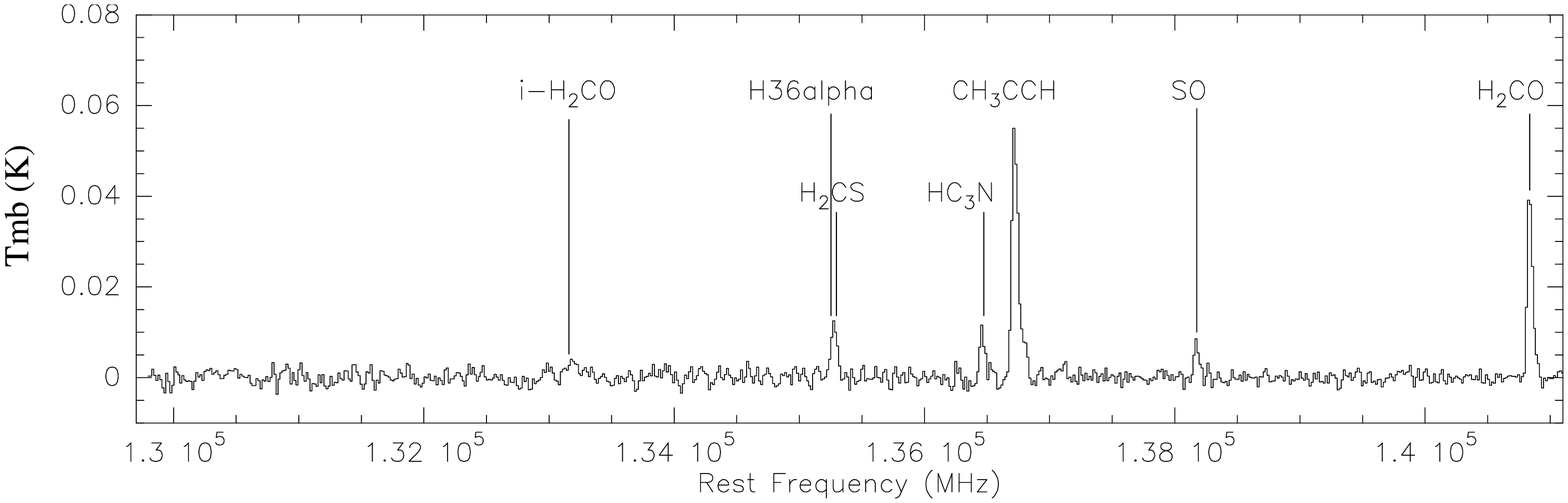}
\includegraphics[angle=0,width=\textwidth]{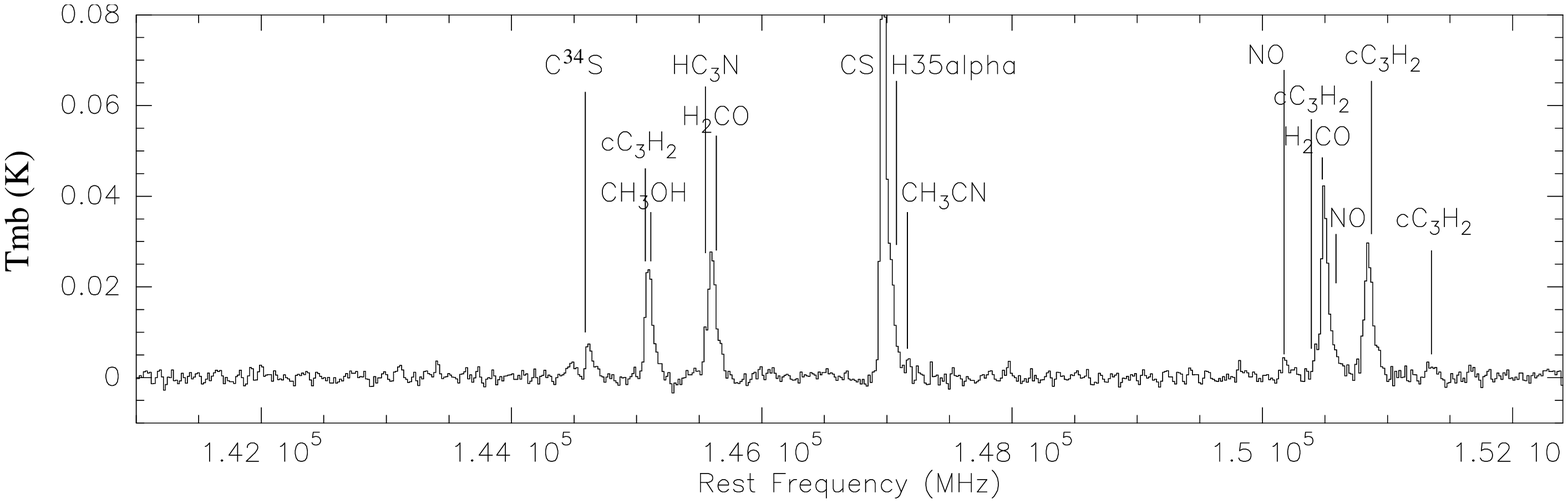}
\includegraphics[angle=0,width=\textwidth]{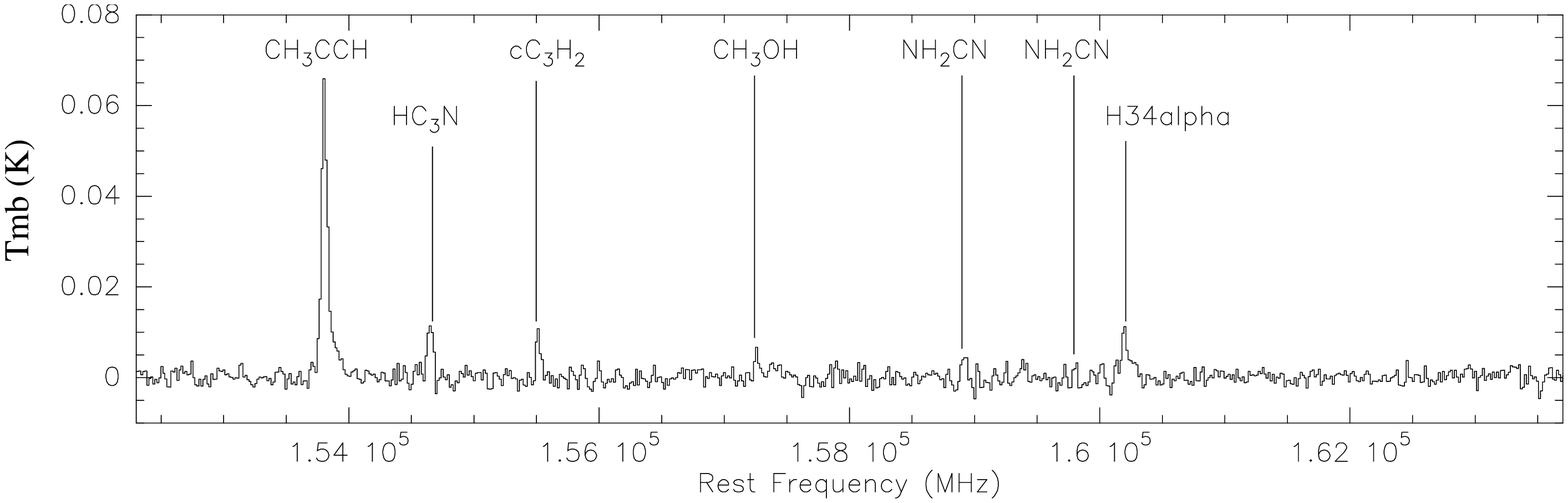}
\includegraphics[angle=0,width=\textwidth]{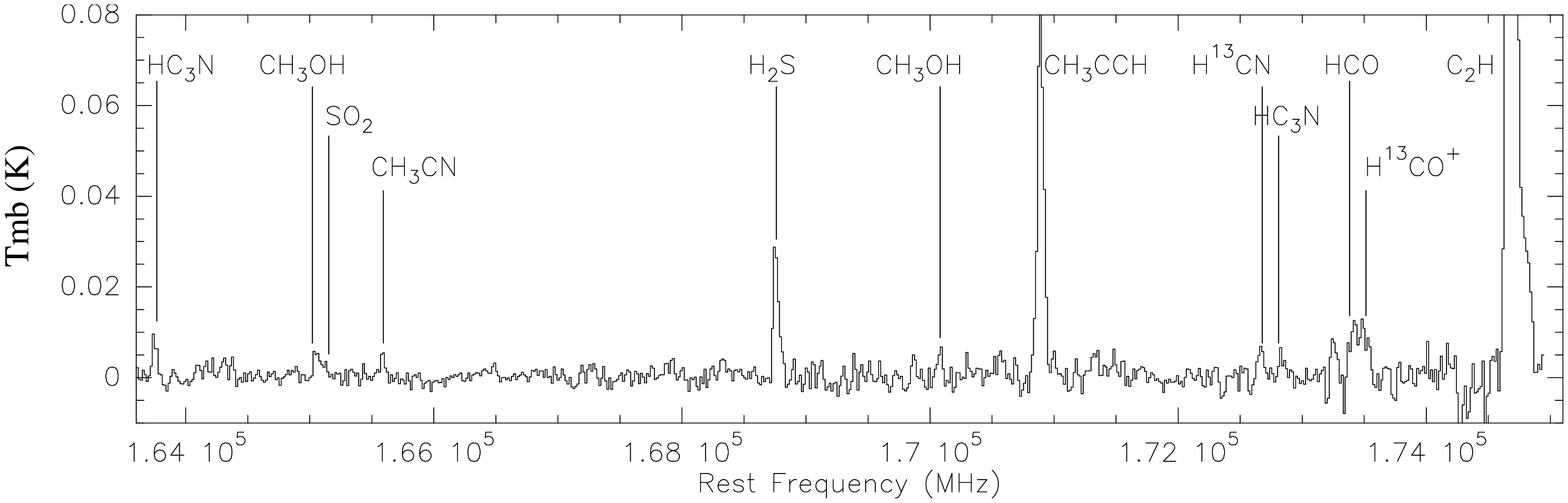}
\caption{M\,82 frequency survey coverage between 129.8\,GHz and 175.0\,GHz and between 241.0\,GHz and 260.0\,GHz. The spectra were smoothed to two channels, corresponding to a channel separation of $\sim$28\,km\,s$^{-1}$ in the 2\,mm atmospheric window and to $\sim$18\,km\,s$^{-1}$ in the 1.3\,mm atmospheric window. Species labeled with {\emph{i-}} come from the image band.}
\label{FigCachos1}
\end{figure*}

\begin{figure*}
\centering
\includegraphics[angle=0,width=0.88\textwidth]{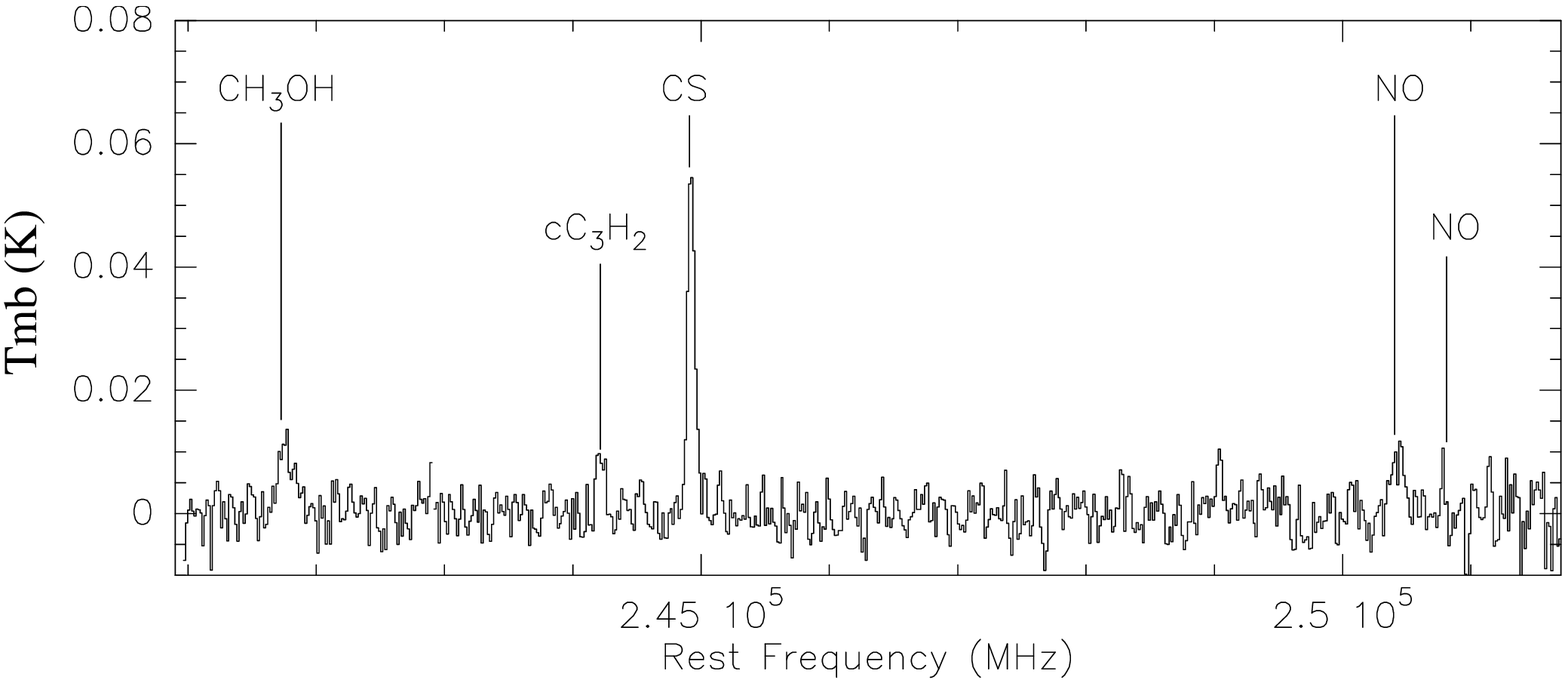}
\includegraphics[angle=0,width=0.88\textwidth]{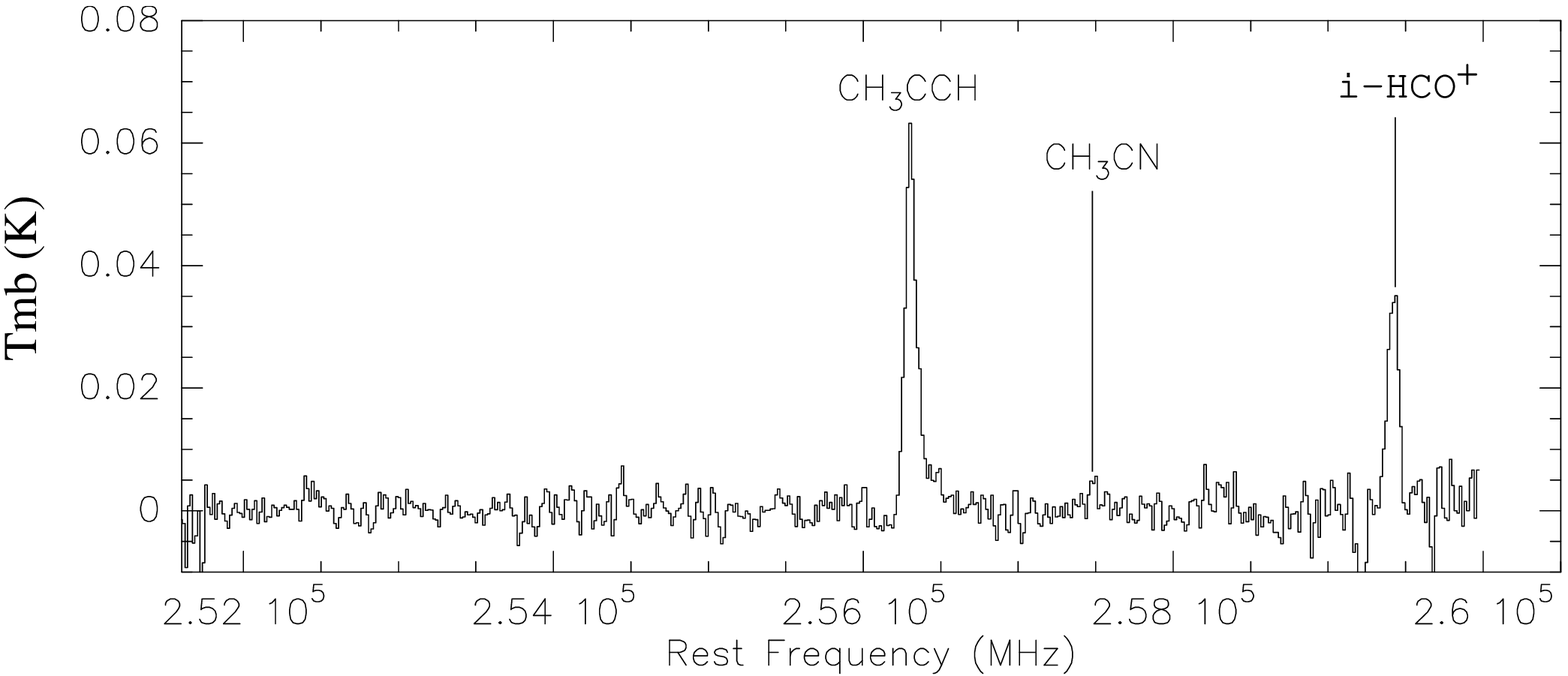}
\addtocounter{figure}{-1}
\caption{Continued}
\label{FigCachos2}
\end{figure*}

\subsection{Rotational temperatures}
\label{Trot}

To accurately determine rotational temperatures from the Boltzmann plots, one needs at least two lines with sufficient energy differences between them. Otherwise, a large range of $T_{\rm rot}$ values satisfies the Boltzmann equation. This is indeed the case for some species for which we detected only one $(J-J')$ transition (e.g. C$_2$H, H$^{13}$CO$^+$, or HCO). In these cases, we fixed $T_{\rm rot}=20\pm10$\,K assuming that this is the average temperature for most of the observed molecules in the NE lobe of M\,82 (see Table~\ref{TableNT}). This assumption might have an important impact on the derived column density for extreme temperatures well above or below this range (i.e. if $T_{\rm rot}>100$\,K or $T_{\rm rot}<5$\,K).

\begin{table*}
\caption{Column densities, rotational temperatures, and fractional abundances of all the observed molecules in the NE lobe of M\,82}
\centering
\begin{tabular}[!h]{lccccccccc} 
\hline
\hline
Molecule\,$^a$ & $N_{\rm mol}$ (cm$^{-2}$) & $T_{\rm rot}^b$ (K) & $N_{\rm mol}/N_{\rm C^{34}S}$ \\

\hline

C$_2$H  & $(1.3\pm0.6)\times10^{15}$	& $20.0\pm10.0$ &  $118.2\pm60.7$ \\ 
\bf{NO}  &	$(1.2\pm0.1)\times10^{15}$	& $8.8\pm4.4$ 	&  $109.1\pm10.7$  \\
CH$_3$CCH\,$^c$ &	$(1.2\pm0.1)\times10^{15}$	& $31.6\pm0.1$ 	&  $109.1\pm10.7$  \\
CS\,(1)\,$^c$ &	$(2.0\pm0.1)\times10^{14}$	& $5.9\pm0.1$ 	&  $22.0\pm1.7$ \\
CS\,(2)\,$^c$ &	$(4.2\pm0.1)\times10^{13}$	& $14.7\pm0.8$ 	&  $\dots$ 	\\
CH$_3$OH &	$(1.2\pm1.1)\times10^{14}$	& $26.2\pm11.7$	& $10.9\pm11.7$ \\
H$_2$CO  & $(1.1\pm0.2)\times10^{14}$	& $25.5\pm3.0$ & $10.0\pm2.7$  \\
\bf{H$_2$S}    &	$(6.1\pm4.5)\times10^{13}$& $20.0\pm10.0$ & $5.5\pm5.2$  \\
c-C$_3$H$_2$ &  $(4.8\pm1.3)\times10^{13}$	& $10.1\pm1.2$ 	& $4.4\pm2.0$  \\
HCO  &$(3.9\pm2.8)\times10^{13}$	& $20.0\pm10.0$ & $3.5\pm3.5$  \\
\bf{SO}$^\dagger$ &	$(3.7\pm0.5)\times10^{13}$ & $20.0\pm10.0$    & $3.4\pm1.2$ \\
\bf{H$_2$CS}  & $(3.1\pm2.5)\times10^{13}$	& $20.0\pm10.0$ & $2.8\pm3.2$  \\
HC$_3$N\,$^c$ &	$(2.5\pm1.2)\times10^{13}$	& $24.2\pm3.9$ & $2.3\pm1.9$ \\ 
\bf{SO$_2$}$^\dagger$ &$(1.2\pm0.6)\times10^{13}$	& $20.0\pm10.0$ & $1.1\pm1.3$ \\
\bf{NH$_2$CN} &	$(1.2\pm1.5)\times10^{13}$	& $80.3\pm52.9$ & $1.1\pm2.2$  \\
C$^{34}$S  & $(1.1\pm0.7)\times10^{13}$	& $20.0\pm10.0$ & $1.0\pm1.0$ \\
\bf{CH$_3$CN} &	$(6.4\pm1.2)\times10^{12}$	& $32.9\pm1.6$ 	& $0.6\pm0.8$  \\
\bf{H$^{13}$CN}$^\dagger$  & $(1.7\pm1.6)\times10^{12}$ & $20.0\pm10.0$ & $0.2\pm0.9$ \\
H$^{13}$CO$^+$  & $(1.3\pm0.8)\times10^{12}$& $20.0\pm10.0$ & $0.1\pm0.8$ \\ 
\hline
C$^{18}$O\,$^d$  & $(3.5\pm1.7)\times10^{16}$ 	&  $20.0\pm10.0$  &  $3181.8\pm1700.7$ \\
CN$\,^e$	&	$(2.0\pm0.5)\times10^{14}$	&	...	&	$18.2\pm5.7$	\\
HCN$\,^e$	&	$(4.0\pm0.5)\times10^{13}$	&	...	&	$3.6\pm1.2$	\\
HNC$\,^f$	&	$2.2\times10^{13}$	&	10.0	&	2.0	\\
CO$^+$\,$^g$  & $(6.6\pm3.4)\times10^{12}$  & $20.0\pm10.0$ &  $0.6\pm1.0$ \\
HOC$^+$\,$^h$  & $(3.9\pm1.9)\times10^{12}$  & $20.0\pm10.0$  &$0.4\pm0.9$ \\

\hline
\end{tabular}
\begin{list}{}{}
\item[\,$^a$]Molecules in boldface are new detections in M\,82.
\item[\,$^b$]Some molecules do not have a wide enough dynamic range in energy to calculate the rotational temperature with precision. For them, we assumed $T_{\rm rot}=20\pm10$\,K (see Sect.~\ref{Trot} and Appendix~\ref{sect.molecules} for details).
\item[\,$^c$] Some extra detections observed by \citet{Aladro10} were used.
\item[] CS shows two components with different column densities and rotational values corresponding to the two fits that result from its Boltzmann diagram. The fractional abundance value is in this case $[N_{\rm CS(1)}+N_{\rm CS(2)}]/N_{\rm C^{34}S}$.
\item[]The molecules shown at the bottom of the table were observed by: $^d$ \citet{Martin09a};  $^e$ \citet{Fuente05} (averaged value for the galaxy disk); $^f$ \citet{ Huttemeister95} (averaged value for the nuclear zone); $^g$ \citet{Fuente06}; $^h$ Aladro et al. in prep.
\item[$^\dagger$] Only upper limits were reported until now: SO, and SO$_2$ by \citet{Petuchowski92}, and  H$^{13}$CN  by \citet{Wild90}.
\end{list}{}{}
\label{TableNT}
\end{table*}

\section{Molecules in M\,82}
\label{Section4}

\subsection{Detected and undetected molecules in the NE molecular lobe}
We detected a total of 69 molecular transitions and identified 18 species, as well as 3 hydrogen recombination lines. While there are no newly detected extragalactic molecules, eight species were detected for the first time in M82. Furthermore, there are fourteen important molecular species which remained undetected. Detected and undetected species are listed in Tables~\ref{TableNT} and ~\ref{TabUpper}, respectively. The line identification was made using the CDMS \citep{Muller01,Muller05}, JPL \citep{Pickett88}, and Lovas \citep{Lovas92} catalogues. 
To confirm whether a suspicious feature is either a weak line or a fake feature, we considered the following criteria: a) those features with line widths much narrower that the typical M82 line widths are unlikely to be weak lines in this galaxy; b) very complex molecules or very rare isotopologues (such as NH$_2$CH$_2$CH$_2$OH, H$_2$NCH$_2$COOHII, $^{35}$ClONO$_2$... etc) are unlikely to be weak lines in M82; c) transitions with high energies of the lower state (usually $>$ 100 cm$^{-1}$) are likely to be too weak to be detected in M82; d) lines with log$_{10}$(integrated intensity) $<$ -6 nm$^2$MHz are likely to be too faint to be detected in M82. After ruling out cases of clear noise spikes, there were some remaining molecules with typical M82 line widths, that were not too complex, of low $E_{\rm low}$, and sufficiently strong. In each case, we examined all the transitions of the species that lie in the survey and checked whether other lines with similar spectroscopic parameters also seemed to be weak lines or not. This scrupulous procedure allowed us to distinguish fake features and weak lines. We note that some of the spectra may contain ripples that were not completely removed because of the low degree of the baseline subtracted, which thus appear as weak lines. We also checked the image sideband for any possible lines. As a consequence of the high rejection, only H$_2$CO\,$(2_{1,2}-1_{1,1})$ and HCO$^+(3-2)$ were detected from the image band at 133.2\,GHz and 259.4\,GHz, respectively.

Gaussian profiles were fitted to all the detected lines. The parameters resulting from those fits are given in Table~\ref{TableGausParameters}. In some cases, emission from the centre of the galaxy was detected by the telescope beam, displaying a bump at lower velocities of the lines. In these cases, we subtracted the bump before fitting the Gaussian profile. In other cases, two or more molecules were blended, thus we calculated the contribution of each line to the total profile and performed a synthetic fitting. The reduction of the spectra and Gaussian profile fitting were performed using the CLASS \footnote{CLASS $http://www.iram.fr/IRAMFR/GILDAS$} and MASSA\footnote{MASSA $http://damir.iem.csic.es/mediawiki-1.12.0/index.php/MASSA\_User's\_Manual$} software packages. Each case is described in Appendix~\ref{sect.molecules}. 

From the LTE analysis, we found that C$_2$H is the most abundant molecule in our M\,82 survey, followed by NO, which is detected for the first time in this galaxy. The $^{13}$C isotopologues H$^{13}$CO$^+$ and H$^{13}$CN have the lowest column densities in the M82 NE molecular lobe (see Table~\ref{TableNT}). We augmented our CH$_3$CCH, CS and HC$_3$N data with other detections observed by \citet{Bayet09b} and \citet{Aladro10} in frequencies that lie outside our frequency coverage. For the whole list of detected molecules, CS is the only one that has two clear different rotational temperatures (of $\sim$6\,K and $\sim$15\,K). ~In addition to CS, NO also has a very low rotational temperature, being $<$10\,K. The molecule NH$_2$CN could have a high rotational temperature (of $\sim$80\,K), that is similar to what is observed in the starburst galaxy NGC\,253 \citep{Martin06b}.

\subsection{Fractional abundances}
\label{FracAbun}
We calculated the fractional abundances of each species with respect to C$^{34}$S (Table~\ref{TableNT}), and the upper limits to the abundances for the undetected species (Table~\ref{TabUpper}). We used the C$^{34}$S\,$(3-2)$ line for this purpose because it is supposed to be significantly more optically thin ($\tau<<1$) than the main isotopologue, which in the worst case could be moderately saturated, with $\tau<2$ \citep{Mauersberger89}. Nevertheless, another abundant optically thin gas tracer, such as C$^{18}$O, could also have been used. We note, however, that C$^{18}$O arises from more extended regions, with critical densities of at least one order of magnitude lower than those traced by C$^{34}$S. Nonetheless, C$^{18}$O is sometimes considered the most effective way to derive H$_2$ column densities. The problem here is that the $^{16}$O/$^{18}$O ratio has not yet been tightly constrained, having only a lower limit of $^{16}$O/$^{18}$O $>$ 350 for M\,82 \citep{Martin10}. As a consequence, the H$_2$ column density cannot be accurately determined. The same argument can be applied to C$^{34}$S, because the sulphur ratio $^{32}$S/$^{34}$S is not well-known either \citep{Martin10}. Nevertheless, we prefer to refer our fractional abundances to the column densities of a dense gas tracer such as C$^{34}$S, because CS and its $^{34}$S isotopologue have similar column densities in M\,82 and NGC\,253 (see Sect.~\ref{sect.comp} for more details).

\subsection{A picture of the nuclear region of M82}
To have a more complete picture of the chemistry in the central region of M82, we compiled from the literature the results of other molecular detections observed towards the central hot-spot and the surrounding molecular ring. As mentioned in Sect~\ref{sect.Obs}, our observations were carried out towards the NE molecular complex, because the HCO observations of \citet{Burillo02} show that the photodissociating radiation is stronger there than in other parts of the nucleus. The emission in the NE molecular lobe is also stronger when observing CS \citep{Bayet08b,Bayet09b}, which is claimed to be a good tracer of dense gas linked to the presence of UV radiation. However, other species such as CO \citep{Mao00}, H$_2$CO \citep{Huttemeister97}, and HNC \citep{Huttemeister95} seem to be more abundant in the SW lobe, while N$_2$H$^+$ and C$_3$H$_2$ \citep{Mauersberger91b} may peak at the very centre. In particular, it was found that a high abundance of CO in combination with a high gas density may lead to a rapid destruction of N$_2$H$^+$ in the molecular ring via the reaction N$_2$H$^+$ + CO $\rightarrow$ N$_2$ + HCO$^+$ \citep{Mauersberger91b,Jorgensen04}.

\begin{table}
\caption{$3\sigma$ upper limits to the column densities and fractional abundances of some important undetected molecules.}
\centering		
\begin{tabular}[!b]{lccccccccc} 

\hline
\hline
Molecule & $N$ (cm$^{-2}$) &  $N/N_{\rm C^{34}S}$ \\

\hline
OCS    &	$\le7.4\times10^{13}$ &	$\le67.0\times10^{-1}$	\\  
HNCO    &	$\le2.5\times10^{13}$	&  $\le23.0\times10^{-1}$ 	 \\ 
C$_2$S   &	$\le1.0\times10^{13}$	& $\le9.0\times10^{-1}$\\ 		
c-C$_3$H   &	$\le9.7\times10^{12}$  & $\le8.8\times10^{-1}$\\ 
C$_2$D   &	$\le9.2\times10^{12}$  & $\le8.4\times10^{-1}$\\ 
CH$_2$NH   &	$\le8.4\times10^{12}$ & $\le7.6\times10^{-1}$	\\ 
$^{13}$CS     &	$\le4.8\times10^{12}$ &	$\le4.4\times10^{-1}$	\\
HOCO$^+$   &	$\le4.2\times10^{12}$	& $\le3.8\times10^{-1}$	\\ 
NS     &	$\le2.3\times10^{12}$ &	$\le2.1\times10^{-1}$\\ 
HN$^{13}$C &	$\le2.3\times10^{12}$ &	$\le2.1\times10^{-1}$\\ 
SiO    &	$\le1.6\times10^{12}$&	$\le1.4\times10^{-1}$\\ 
DCN    &	$\le1.2\times10^{12}$&	$\le1.1\times10^{-1}$\\
DNC	&	$\le7.0\times10^{11}$	&  $\le3.4\times10^{-2}$	\\
N$_2$D$^+$&	$\le5.3\times10^{11}$	&  $\le4.8\times10^{-2}$	\\ 
\hline
\end{tabular}

\begin{list}{}{}
\item[]To calculate the upper limits to the column densities, we assumed $T_{\rm rot}=20$\,K  and $\Delta\rm v=100\,km\,s^{-1}$ for all the molecules.
\item[]Fractional abundances are compared to those derived for NGC\,253 in Fig.~\ref{FigCompAbundLimits}.
\end{list}{}{}
\label{TabUpper}
\end{table}

Each molecule may arise from gas that can be at different temperatures. There is a notable difference between the temperature derived from two of the most important thermometers, H$_2$CO and NH$_3$. \citet{Muhle07} carried out a multi-transition analysis of formaldehyde in the NE and SW molecular lobes. They derived densities of $7\times10^3$\,cm$^{-3}$ and high kinetic temperatures around 200\,K, and found that H$_2$CO seems to be co-spatial with CO. The relatively low densities obtained for both species are quite similar. However, the study performed by \citet{Weiss01b} towards the SW molecular lobe using ammonia indicates that the NH$_3$ emitting gas is rather cold, with kinetic temperatures of $\sim60$\,K. The fractional abundance of NH$_3$ is extremely low in M82, because this molecule is easily destroyed by the UV radiation. If we were to assume that the ammonia properties in the NE lobe are similar to those in the SW, it could be said that most of the observed molecular gas can be more accurately represented by H$_2$CO rather by NH$_3$. On the other hand, the low density determined from the H$_2$CO in M82 could lead to highly subthermal excitation. Furthermore, we expect that at least part of the detected molecular emission arises from well quite-shielded molecular regions, which also produce NH$_3$ emission, and  should be relatively cool. Thus, the kinetic temperature should be considered in each single case to see whether the properties and spatial distribution of a given molecule are more similar to those of formaldehyde or ammonia.

\section{Comparison of the M\,82 and NGC\,253 chemistries}
\label{sect.comp}
 
The galaxies M\,82 and NGC\,253 share several characteristics that have ensured that they are among the most studied of extragalactic objects. Apart from their luminosities ($L_{\rm IR}^{\rm NGC\,253}=2.1\times10^{10}\,L_\odot$, \citealt{Vaucou91}; $L_{\rm{IR}}^{\rm M\,82}=3\times10^{10}\rm{L_\odot}$, \citealt{Rice88,Telesco88}), they lie at a similar distance (D$_{\rm NGC\,253}$=3.9\,Mpc, D$_{\rm M\,82}$ = 3.6\,Mpc, \citealt{Freedman94}), so their morphology, physical processes, and chemistry can be studied in detail with the same spatial resolution. Furthermore, both nuclear regions host very active star-forming regions (SFR$_{\rm NGC\,253}\sim$3.6\,M$_\odot$\,yr$^{-1}$, SFR$_{\rm M\,82}\sim9\,\rm{M_\odot}\,\rm{yr}^{-1}$, \citealt{Strickland04}). 
 
\begin{figure*}
\begin{center}
\includegraphics[width=\textwidth]{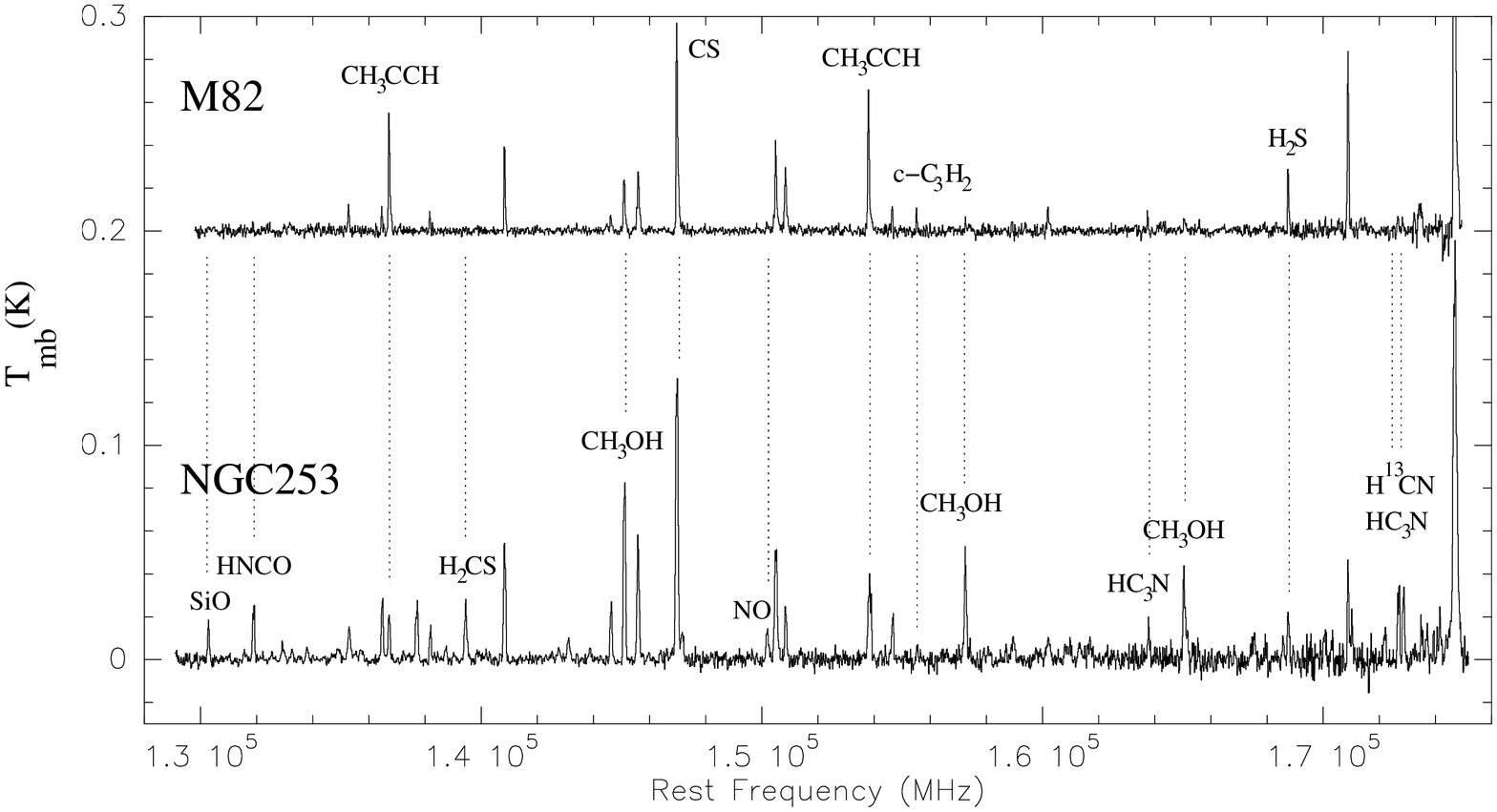}
\caption{Comparison of the M\,82 and NGC\,253 surveys at 2\,mm wavelengths. Both spectra have been smoothed to a velocity resolution of $\sim$\,28\,kms$^{-1}$. At first glance, NGC\,253 shows a greater chemical complexity than M\,82, with more lines and molecular species.}
\label{FigCompM82NGC253}
\end{center}
\end{figure*}

\citet{Martin06b} carried out a frequency survey towards the central region of NGC\,253 covering the same frequency range in the 2\,mm atmospheric window as the M\,82 survey presented here. Up to 25 species were identified in NGC\,253. The combination of both surveys allows us, for the first time, to perform a detailed and complete comparison of the chemical composition in these prototypical starbursts. Figure~\ref{FigCompM82NGC253} shows the comparison of the 2\,mm scans towards both galaxies. In Fig.~\ref{FigCompAbund}, we compare the fractional abundances of the 18 species detected in both line surveys. We also included other important molecules from the literature, such as C$^{18}$O \citep{Martin09a,Martin09b}, CO$^+$ \citep{Martin09b,Fuente06}, and HOC$^+$ (\citealt{Martin09b}, Aladro et al. in prep.) since they do not have transitions in the covered frequency range. Moreover, we also show in Fig.~\ref{FigCompAbundLimits} a comparison of the fractional abundances of 12 other important molecules that were not detected in our M\,82 survey, but towards NGC\,253. Therefore, we plot $3\sigma$ upper limits to their abundances. In addition, Fig.~\ref{FigTempM82NGC253} compares the rotational temperatures of several molecules in M\,82 and NGC\,253.

Next, we discuss in detail all the species, emphasizing those that allow us more clearly differentiate between the M\,82 and NGC\,253 chemistries. 

\begin{itemize}

\item {\bf C$_2$H}. The ethynyl radical may be strongly related to the C{\sc i} abundance and its formation can be explained in terms of gas-phase chemistry \citep{Turner99}. The main formation path of C$_2$H is via electron recombination reactions with the C$_2$H$_2^+$ and C$_2$H$_3^+$ ions \citep{Watson74,Wootten80,Mul80}. It has been claimed to be a good photon-dominated region tracer \citep{Huggins84,Henkel88,Meier05,Fuente05,Bayet09b}. The molecule C$_2$H is a factor $\sim3$ more abundant in M\,82 than in NGC\,253. Furthermore, it is the brightest spectral feature in both galaxy surveys. 
 
Although this molecule was detected in a large number of Galactic sources, it went almost unnoticed in external galaxies, except for the M\,82 and NGC\,253 detections (see also \citealt{Martin10}), up to now it has only been observed towards one additional galaxy, namely IC\,342. There it was found in the central $50-100$\,pc region, where radiation fields are strong \citep{Meier05}.

\item{\bf CS}. Carbon monosulfide is one of the most well-studied dense gas tracers \citep{Bayet08a} in extragalactic sources, and in particular starburst galaxies  \citep{Henkel85,Mauersberger89,Mauersberger89b,Bayet08b,Bayet09b,Aladro10}. The main chemical reactions that lead to the CS formation are detailed in \citet{Pineau86} and \citet{Drdla89}. A high CS abundance is predicted for slightly shielded regions \citep{Sternberg95}. The CS column densities and fractional abundances with respect to C$^{34}$S are almost the same in both M\,82 and NGC\,253, which is  unsurprising if CS saturation is  insignificant (or comparable), and the $^{32}$S/$^{34}$S isotope ratios (both sulphur nuclei are formed by oxygen burning in massive stars) are similar (see also \citealt{Requena06} and \citealt{Martin09a}).

It can be seen in Fig.~\ref{FigTempM82NGC253} that CS has up to three $T_{\rm rot}$ components for NGC\,253, while M\,82 only shows two components  (Fig.~\ref{FigDR}, Table~\ref{TableNT}). The highest $T_{\rm rot}$ value of NGC\,253 is well above the highest one in M\,82. This could be explained if the gas in NGC\,253 had a higher kinetic temperature. It seems that in these two galaxies, the spatial distribution of CS is similar to that of ammonia, or at least, is clumpier than those of H$_2$CO and CO \citep{Aladro10}. If so, the kinetic temperature in M\,82 would be around 60\,K, while that of NGC\,253 would be between 100 and 120\,K. In that case, NGC\,253 does not necessarily need to have higher $\rm H_2$ densities than M\,82, as seen by the LVG modelling of the CS emission towards these two galaxies by \citet{Aladro10}.

As mentioned in Sect.~\ref{FracAbun}, C$^{34}$S is strong enough to be clearly detected in both galaxies and is observed to be similarly abundant. On the other hand, although $^{13}$CS was detected in NGC\,253, it is fairly faint in M\,82, where we only have an upper limit. Finally, the double isotopologue $^{13}$C$^{34}$S is too faint to be detected in both galaxies. A comparison of the CS\,$(3-2)$ and C$^{34}$S\,$(3-2)$ integrated intensities indicates a sulphur isotopic ratio of $^{32}$S\,/\,$^{34}$S$\,\sim$\,12. This value may be higher, since CS could be moderately optically thick in M\,82 \citep{Mauersberger89}. Using the upper limit to $^{13}$CS (see Table~\ref{TabUpper}), we also calculated a lower limit to the $^{12}\rm C/^{13}\rm C$ carbon isotopic ratio. We obtained $^{12}\rm C\,/\,^{13}\rm C>50$. This limit is lower than the one obtained by \citet{Martin10} using the molecule CCH and its carbon isotopologues ($^{12}\rm C/^{13}\rm C>138$).

\begin{figure*}
\begin{center}
\includegraphics[angle=0,width=\textwidth]{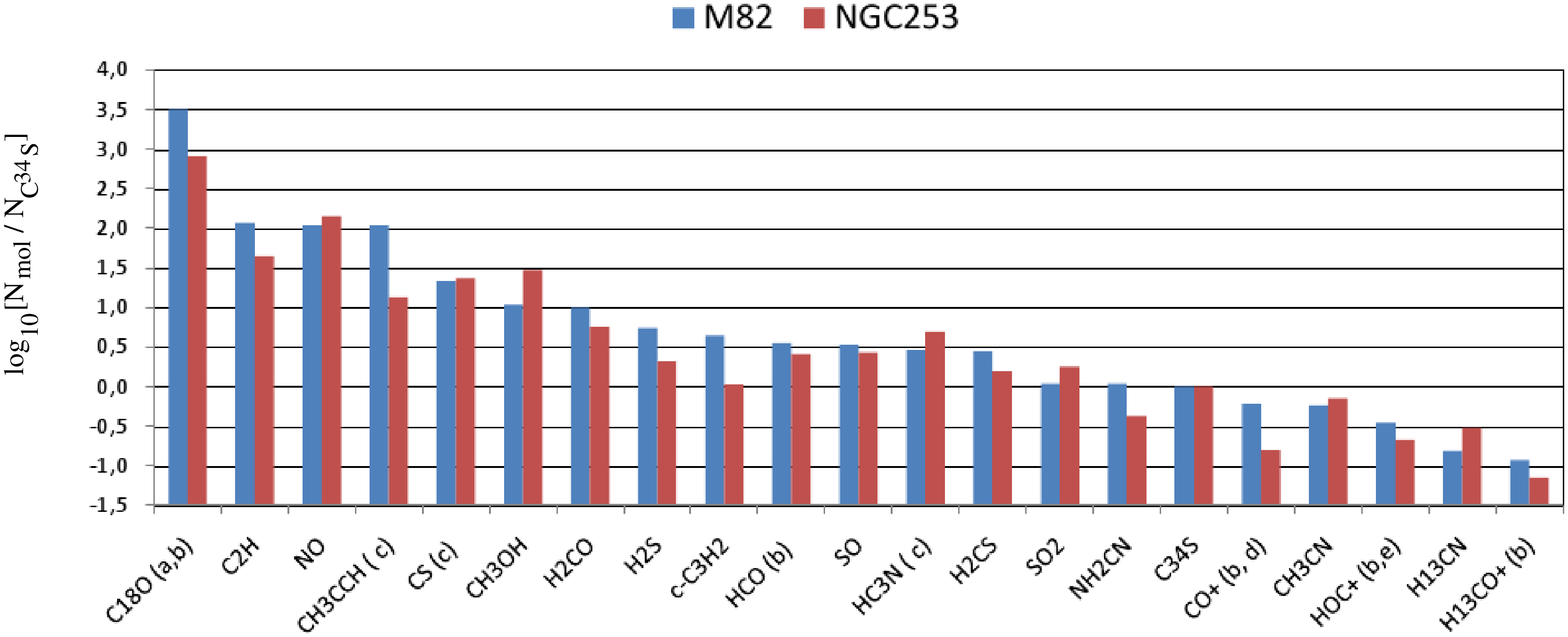}
\caption{Comparison of the M\,82 and NGC\,253 fractional abundances of different molecules with respect to the C$^{34}$S\,(3-2) abundances. All the fractional abundances have been calculated from the millimeter frequency surveys of \citet{Martin06b} and this paper except for: (a) C$^{18}$O: \citet{Martin09a}; (b) C$^{18}$O, HCO, HOC$^+$, and H$^{13}$CO$^+$: \citet{Martin09b}; (c) CH$_3$CCH, CS and HC$_3$N: \citet{Aladro10}; (d) CO$^+$: \citet{Fuente06}; (e) HOC$^+$: Aladro et al. in prep.}
\label{FigCompAbund}
\end{center}
\end{figure*}

\item {\bf c-C$_3$H$_2$}. As for CS, cyclopropenylidene is also formed by means of gas phase ion-molecule reactions, from the dissociative recombination of the stable cyclopropenyl cation, C$_3$H$_3^+$ and its cyclic and linear isomers, c-C$_3$H$_3^+$  and l-C$_3$H$_3^+$ \citep{Thaddeus85,Adams87}. Thus, the destruction of some species by UV fields and the subsequent creation of ions would favour an enhancement of the  c-C$_3$H$_2$ abundance in M\,82. Our M\,82 data indeed indicate a fractional abundance four times higher than in NGC\,253. The rotational temperature derived from the Boltzmann diagram of M\,82 is $10.1\pm1.2$\,K (similar to the temperatures obtained in NGC\,253). This seems to indicate that cyclopropenylidene arises from the envelopes of the molecular clouds.

\item {\bf H$^{13}$CO$^+$, H$^{13}$CN}. These two species are the only $^{13}$C isotopologues detected in our survey, and have the lowest abundances among all the molecules detected in this study. The ratio of their main carbon isotopologues, HCO$^+$/HCN, has been studied to differentiate between the physical processes occurring in starburst galaxies and AGN environments, and has been found to be lower in the latter \citep{Kohno01,Imanishi04,Imanishi09,Krips08}. The HCO$^+$/HCN ratio also reflects how evolved a starburst regions is, being higher in evolved starbursts (where there are strong ionization effects) than in young starbursts (that are more dominated by shocks). This ratio can also be calculated using the $^{13}$C isotopologues, with the advantage that the $^{13}$C lines are likely to be optically thinner, thus avoiding line opacity effects. From our data, we obtain H$^{13}$CO$^+(2-1)$\,/\,H$^{13}$CN$(2-1)=1.9$. This is significantly higher than  HCO$^+(1-0)$\,/\,HCN$(1-0)=1.4$ and HCO$^+(3-2)$\,/\,HCN$(3-2)=1.2$, derived by \citet{Krips08} using the main isotopologues for the same region in M\,82. However, the difference in the ratios HCO$^+$/HCN and H$^{13}$CO$^+$/H$^{13}$CN may also be due to opacity effects since the HCN line could be less saturated (because of the hyperfine structure) than the HCO$^+$ lines. On the other hand, our  H$^{13}$CO$^+(2-1)$\,/\,H$^{13}$CN$(2-1)$ ratio for M\,82 is a factor of $\sim5$ higher than the one found for the same transitions in NGC\,253 (0.4, \citealt{Martin06b}). This difference between M\,82 and NGC\,253 is consistent with the idea that evolved starbursts have higher HCO$^+$\,/\,HCN ratios.

\item {\bf HCO, HOC$^+$, CO$^+$}. These molecules are claimed to be good PDR tracers, with their abundances being enhanced in zones where the chemistry is strongly influenced by UV and X-ray radiation \citep{Burillo02,Savage04,Fuente05,Fuente08,Gerin09,Martin09b}. Thus, these species are formed through ion-molecule reactions in gas phase processes, although the chemical path of HCO is less clear and could also involve neutral-neutral reactions \citep{Watson75,Hollis83,Spaans07}. The HCO-to-H$^{13}$CO$^+$ ratio has been found to be a good indicator of UV fields in the before mentioned zones. From our data, we find that $N_{\rm HCO(2-1)}$\,/\,$N_{\rm H^{13}CO^+(2-1)}$\,=\,30. This value is similar to those found in NGC\,253 \citep{Martin09b}, and Orion \citep{Schilke01}. However, this ratio is a factor of ten higher than the one found by \citet{Burillo02} for the whole disk of M\,82 ($N_{\rm HCO(1-0)}$\,/\,$N_{\rm H^{13}CO^+(1-0)}\sim3.6$). On the other hand, the ratios of CO$^+$ and HOC$^+$ with respect to HCO$^+$ calculated towards both galaxies are similar \citep{Martin09b}. 

To complete our sample of molecules, we included the HOC$^+(1-0)$ and CO$^+({5/2}_2-{3/2}_1)$ lines from \citet{Martin09b}, \citet{Fuente06}, and Aladro et al. in preparation (see Table~\ref{TableNT} and Fig.~\ref{FigCompAbund} for details). As expected, both molecules are more abundant in M\,82 than in NGC\,253. This is especially true for CO$^+$, whose fractional abundance is almost four times higher. The large amount of this short-lived reactive ion is clearly indicative of the strong PDRs towards the NE molecular lobe of M\,82, as shown by \citet{Fuente06}.

\item {\bf CH$_3$OH} and {\bf HC$_3$N}. Methanol and cyanoacetylene can be efficiently formed on grain mantles rather than by means of gas-phase reactions \citep{Hartquist95,Hidaka04}, and are claimed to be good tracers of dense gas \citep{Rodriguez98,Bayet08a,Aladro10}. Methanol can also trace shocks when it is released from the grain mantles to the gas phase through grain sputtering in shocks. Both species are supposed to be easily dissociated in the external and less dense layers of the molecular clouds. At the same time, their abundances increase in the molecular cloud cores, where the gas is well-protected from the UV radiation. Both CH$_3$OH and HC$_3$N have fractional abundances 2-3 times higher in NGC\,253 than in M\,82. This is consistent with the results of \citet{Snell11}. 

The temperature and colunm density that we obtained for methanol in M\,82 are similar to the ones found by \citet{Martin06a} for the densest methanol component detected in the NE molecular lobe of M\,82. However, we do not find evidence of the colder and less dense gas component reported by \citet{Martin06a} (with $N_{\rm CH_3OH}=2.7\times10^{13}$\,cm$^{-2}$ and $T_{\rm rot}=5.0\pm0.1$\,K), since this component is traced by a lower excitation transition at  3\,mm wavelengths

Methanol generally arises from cold gas in galaxies \citep{Martin06a,Martin06b}, which is in good agreement with the NGC\,253 rotational temperature seen in Fig.~\ref{FigTempM82NGC253}. However, the $T_{\rm rot}$ in M\,82 is warmer ($\sim26$\,K), leading to a difference in rotational temperatures between both galaxies of $\sim15$\,K. 

On the other hand, cyanoacetylene usually traces very excited and warm gas \citep{Aalto07}. While this molecule identifies a constant temperature distribution in M\,82 ($T_{\rm rot}\sim24$\,K), it shows up to three density components with different rotational temperatures in NGC\,253 (the component belonging to the higher excitation transitions has a rotational temperature of $\sim 73$\,K). This is due to the different density structure of the molecular clouds in both galaxies \citep{Aladro10}. The molecule  HC$_3$N indeed displays the largest difference in $T_{\rm rot}$ between M\,82 and NGC\,253 ($\Delta T_{\rm rot} \sim50$\,K, for the warmest NGC\,253 component, Fig.~\ref{FigTempM82NGC253}). That HC$_3$N has a very warm $T_{\rm rot}$ in NGC\,253 (73\,K) seems to be related to the higher kinetic temperature of this galaxy, since the averaged n$_{\rm{H_2}}$ densities are lower in NGC\,253 than in M\,82 \citep{Aladro10}. 

\begin{figure}
\begin{center}
\includegraphics[angle=0,width=0.54\textwidth]{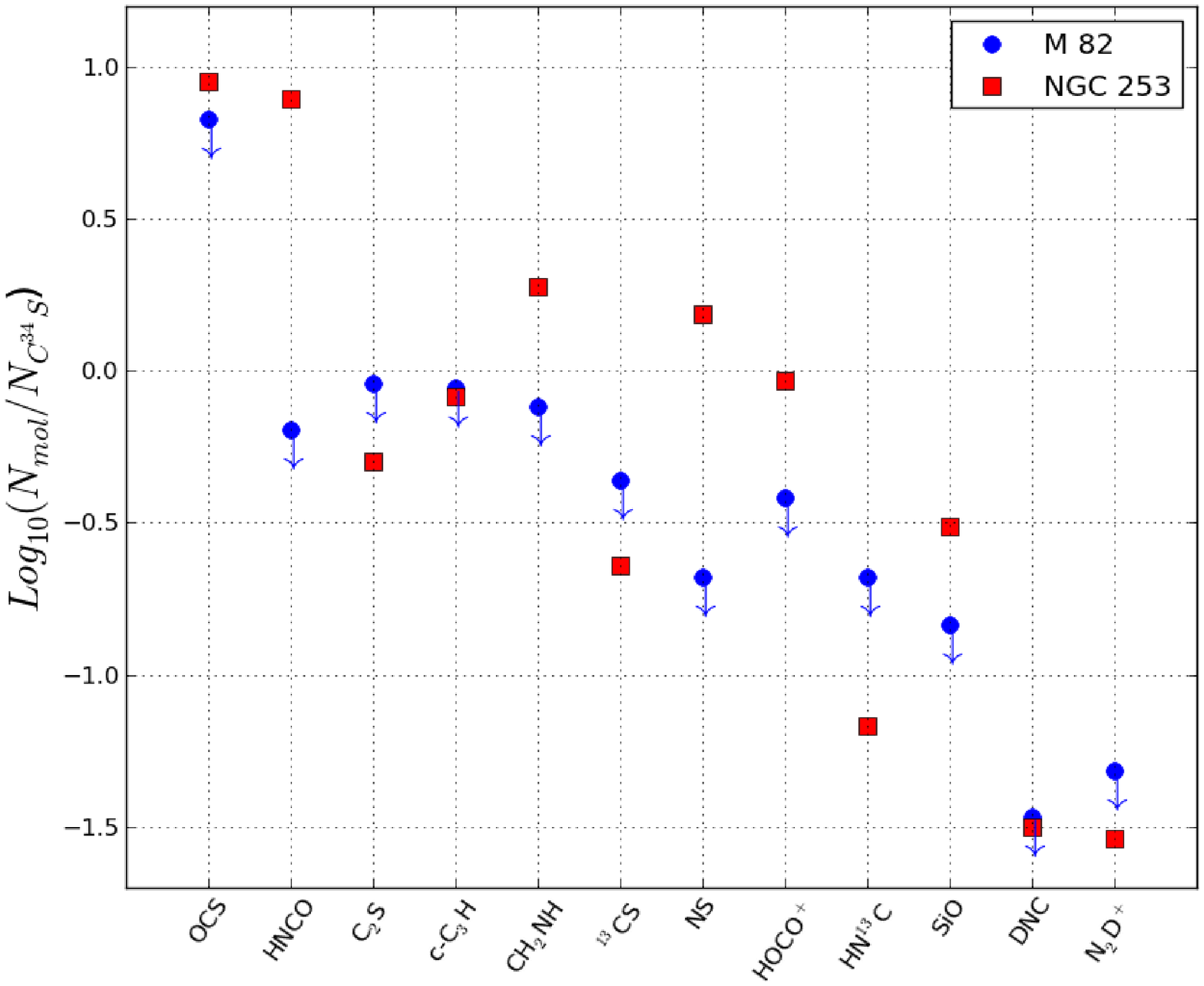}
\caption{Fractional abundance limits of twelve undetected molecules in our M\,82 survey compared to the NGC\,253 abundances taken from the frequency survey done by \citet{Martin06b}. HNCO abundances were taken from \citet{Martin09a}.}
\label{FigCompAbundLimits}
\end{center}
\end{figure}

The [c-C$_3$H$_2$/HC$_3$N] ratio appears to indicate the evolutionary stage of the bursts of star formation \citep{Fuente05}. Since c-C$_3$H$_2$ is enhanced in PDRs, and HC$_3$N is instead destroyed in these regions, this ratio should be higher in M\,82 than in NGC\,253. Comparing our results with those of \citet{Martin06b}, we find a difference amounting to one order of magnitude with $N_{\rm c-\rm C_3\rm H_2}\,/\,N_{\rm HC_3\rm N}=2.0$ and 0.2 for M\,82 and NGC\,253, respectively.

\item {\bf CH$_3$CN}. Methyl cyanide can be formed via gas-phase, where CH$_3^+$ and HCN are precursors \citep{Mauersberger91b,Millar97}. These molecules CH$_3$CN and CH$_3$CCH are good tracers of dense gas that have been used as thermometers \citep{Churchwell83,Guesten85}. The difference in their dipole moments (0.75 Debye for CH$_3$CCH and 3.9 Debye for CH$_3$CN) can lead to differences in the $T_{\rm rot}$, because molecules with low dipole moment are expected to trace temperatures closer to the kinetic temperature. However, we find similar rotational temperatures for both molecules in M\,82 ($T_{\rm rot}=32.9\pm1.6$\,K for CH$_3$CN and $T_{\rm rot}=31.6\pm0.1$\,K for CH$_3$CCH). The expected behaviour is observed in NGC\,253, where CH$_3$CN shows a lower $T_{\rm rot}$ than CH$_3$CCH ($9.6\pm0.7$\,K and $63.0\pm20.0$\,K respectively, \citealt{Martin06b}). Thus, the relationship between the dipole moments and $T_{\rm rot}$ is unclear. In any case, CH$_3$CN arises from a more clumpy gas, possibly from the cores of the molecular clouds, while CH$_3$CCH, more abundant and extended, arises from intermediate density regions \citep{Aladro10}. Furthermore, CH$_3$CN seems to be tracing gas with different properties in both galaxies: the molecular cloud cores of M\,82 traced by this species seem to be warmer than those of NGC\,253 (Fig.~\ref{FigTempM82NGC253}). 

\item {\bf Sulphur-bearing species}. In general, sulphur is believed to freeze onto grain mantles, and experience a later evaporation by shocks that leads to the subsequent formation of several sulphur species, such as OCS, SO, SO$_2$, H$_2$S, or H$_2$CS \citep{Scalo80,Millar90}. However, sulphur might  frequently alternate between a grain-phase and a gas-phase because of the volatile and fragile nature of some of the molecular compounds it forms \citep{Minh91,Omont93}. The aforementioned species are not only shock tracers but also tracers of high-mass star formation in starburst galaxies such as M\,82, as proved by the models of \citet{Bayet08a}.

We have found that SO, SO$_2$, and H$_2$CS have similar abundances in M\,82 and NGC\,253. This indicates that exist shocks in the NE molecular lobe of the galaxy. We know that M\,82 is not a pure-PDR but it also contains significant amounts of dense UV-protected gas \citep{Martin06a} affected by shocks, as  reported by \citet{Burillo01} using SiO observations. The low SiO abundance in M\,82 (in comparison with that of NGC\,253) was  firstly measured by \citet{Mauersberger91a}, \citet{Sage95}, and \citet{Burillo00}. Similarly, a lower OCS abundance was reported in M82 \citep{Mauersberger95}, which we  confirm here by its non-detection in our line survey.

\item {\bf H$_2$CO}. Formaldehyde can be efficiently formed on the surface of icy grains via the hydrogenation of CO \citep{Charnley97,Watanabe02}. It is claimed to be a tracer of dense gas \citep{Bayet08a}. In those occasions when it is released into the gas-phase by shocks (thus also behaving as a shock tracer), it has abundances similar to those of other molecules such as methanol or SiO \citep{Bachiller97,Rodriguez2010}. The H$_2$CO fractional abundances (relative to C$^{34}$S) of M\,82 and NGC\,253 are similar, although that of M\,82 might be the consequence of photo-evaporation rather than shocks.

The rotational temperatures of H$_2$CO are similar in NGC\,253 and M\,82 ($\sim26$\, and $31$\,K respectively, see Fig.~\ref{FigTempM82NGC253}).

\begin{figure}
\begin{center}
\includegraphics[width=0.54\textwidth]{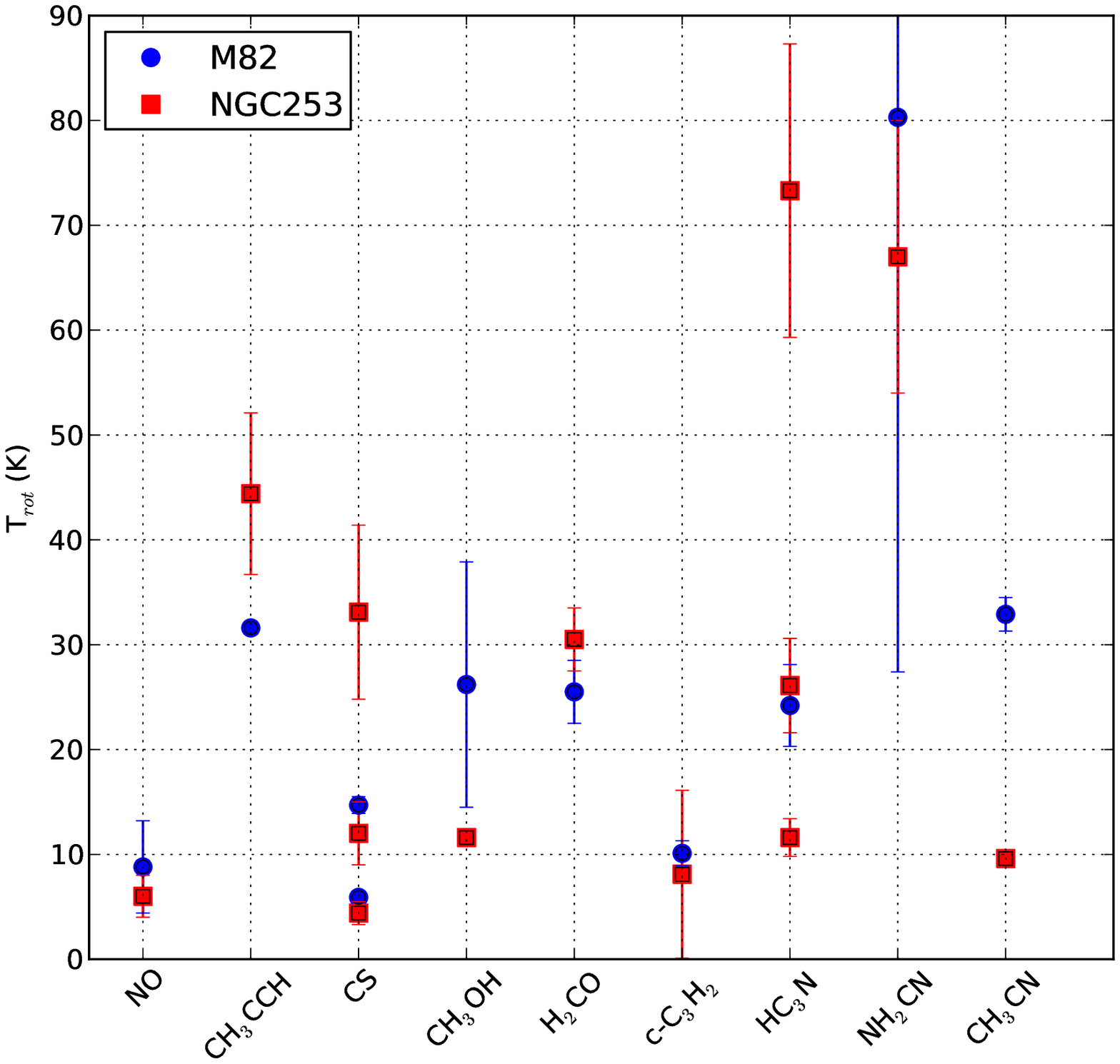}
\caption{Comparison of the M\,82 and NGC\,253 rotational temperatures for several molecules.}
\label{FigTempM82NGC253}
\end{center}
\end{figure}

\item {\bf HNCO}. Chemical models include dust-grain surface reactions and gas-phase reactions for explaining the formation of isocyanic acid (HNCO), because gas-phase reactions alone seem to be inefficient \citep{Caselli93,Garrod08,Martin09b}. The molecule HNCO is supposed to be released through grain mantle evaporation by shock waves and/or by neutral-neutral reactions in the gas phase involving CN and O$_2$ \citep{Rodriguez2010}.This molecule has emerged as one of the most likely promoting star formation in the molecular clouds of both the Milky Way and external galaxies \citep{Martin09a,Rodriguez2010}. Its abundance varies by nearly two orders of magnitude among galaxies dominated by different physical processes \citep{Martin09a}. Our data support this idea, because this molecule is not detected in our survey, while there is a clear detection in NGC\,253. The molecule HNCO has never been detected towards M\,82 (see also \citealt{Nguyen91}), presumably because this molecule is very sensitive to photodissociation \citep{Martin09b}.

\item{\bf c-C$_3$H}. Although the production mechanism of cyclopropynylidyne is not well-known, it has been proposed that it has the same ion precursor as C$_3$H$_2$, namely C$_3$H$_3^+$, in dark clouds \citep{Adams87,Mangum90}. In PDRs, the C$_3$H$_2$ abundance is seen to be higher than that of c-C$_3$H. As an example, in the Horsehead nebula the c-C$_3$H fractional abundance is one order of magnitude lower than the former \citep{Teyssier04}. Though c-C$_3$H could be related to local PDRs, in M\,82 it is too faint to be detected. We find the ratios C$_2$H/c-C$_3$H\,$>$\,100 and c-C$_3$H$_2$/c-C$_3$H\,$>$\,5. The upper limit to the abundance derived from our non-detection is still similar to the value derived from NGC\,253.

\item {\bf NS}. We describe this sulphur-bearing molecule in more detail because it shows the largest fractional abundance contrast between M\,82 and NGC\,253 after HNCO and CH$_3$CCH. This molecule was detected for the first time outside our Galaxy by \citet{Martin05} in NGC\,253, and since then no other extragalactic detection has been reported. It was claimed that the NS chemistry is driven by the large-scale shocks that dominate the NGC\,253 galaxy nucleus. On the other hand, the NS\,/\,CS ratio was proposed as an indicator of the presence of shocks in hot cores \citep{Viti01}. We obtain NS\,/\,CS\,$<$\,0.01 and NS\,/\,CS\,=\,0.1 for M\,82 and NGC\,253 respectively, using the data from \citet{Martin06b}. This difference of at least one order of magnitude indicates that UV fields affect the chemical abundance of NS, and indicates that this ratio is also a good indicator of shocks in galaxies. 

\item {\bf HOCO$^+$}. Protonated carbon dioxide (HOCO$^+$) is thought to be formed in the gas phase from desorbed CO$_2$ \citep{Turner99}. In our Galaxy, it was observed towards translucent clouds and the Galactic centre, being especially abundant in Sgr\,A and Sgr\,B2 \citep{Minh88,Deguchi06}. The enhancement of this molecule in these regions was linked to shock activities that release the CO$_2$ frozen onto grains into the gas-phase. Outside the Milky Way, NGC\,253 is the only galaxy towards which this molecule has  so far been detected. Our upper limit in M\,82 shows that it is at least 2.5 times less abundant than in NGC\,253. This underabundance might be due to a lower influence of shocks in M\,82.

\item {\bf CH$_3$CCH}. After HNCO, propyne is the molecule that has the largest difference between the  abundances of NGC\,253 and M\,82, which are eight times higher in the latter galaxy. This difference in the abundances was also found by \citet{Snell11}, who detected CH$_3$CCH in M\,82 but not NGC\,253. The column density that we derive in M\,82 ($1.2\times10^{15}$\,cm$^{-2}$) is the highest ever measured for an extragalactic source. This results in a fractional abundance CH$_3$CCH/H$_2=1.5\times10^{-8}$, that is similar to the abundance found towards Orion (between $\sim10^{-8}$ and $\sim10^{-9}$, \citealt{Churchwell83,Sutton85}). The most plausible explanation of the high abundance of this molecule in M\,82 is that it was formed in the gas phase, in which the destruction of propyne by UV fields could be balanced by efficient formation by means of ion-molecule or neutral-neutral reactions \citep{Millar84,Bisschop07}. In addition, propyne is observed in dense regions that are well-shielded from the dissociating radiation. Fig.~\ref{FigTempM82NGC253} shows that the CH$_3$CCH rotational temperature is more than 10\,K higher in NGC\,253 than in M\,82. This is due to a possible higher kinetic temperature in NGC\,253 \citep{Mauersberger03}, as well as a higher H$_2$ density in the molecular clouds of NGC\,253 traced by CH$_3$CCH \citep{Aladro10}.

\item {\bf NO}. Nitric oxide is detected for the first time in M\,82. It is the main precursor of the N/O chemical network, and its abundance is directly related to the nitrogen and oxygen abundances by the reactions NH $+$ O $\rightarrow$ NO $+$ H \citep{Halfen01} and N $+$ OH $\rightarrow$ NO $+$ H \citep{Lique09,Bergeat11}. 

We note that the nitric oxide abundances are almost the same in M\,82 and NGC\,253. These values indicate that it is not strongly affected by UV fields or shocks, as in the case of CS. On the other hand,  NO is likely to be widespread across the nuclear regions. The NO dipole moment is quite similar to that of CO and it is therefore likely to trace the overall gas. The low rotational temperatures of NO in both galaxies ($T_{\rm rot}=8.8\pm4.4$\,K for M\,82 and $6\pm2$\,K for NGC\,253)  also indicate that it  possibly originates in the external and colder layers of the molecular clouds.

\item{\bf NH$_2$CN}. The formation process of cyanamide is not well-known, since this molecule was observed only towards a few sources. Its column density in M\,82 is consistent with that found in Sgr\,B2  by \citet{Cummins86}  ($N_{\rm NH_2CN}=2\times10^{13}$\,cm$^{-2}$) and \citet{Turner75} ($N_{\rm NH_2CN}\sim10^{14}$\,cm$^{-2}$). However, cyanamide is so faint in M\,82 that we only have tentative detections. As a consequence, its rotational temperature and column density have very large errors, so these results should be taken with caution. The molecule NH$_2$CN was not detected in other local massive star-forming regions such as Orion \citep{Sutton85}. \citet{Turner75} concluded that this molecule probably arises from high density regions with n\,$\gtrsim$\,10$^5$cm$^{-3}$. In external galaxies, cyanamide was also detected towards NGC\,253 \citep{Martin06b}. In contrast to the Sgr\,B2 molecular cloud (OH position), where the excitation temperature of this molecule was found to be relatively low ($T_{\rm rot}=15\pm3$\,K, \citealt{Cummins86}), cyanamide shows a very high rotational temperature in NGC\,253 ($T_{\rm rot}=67\pm13$\,K). The latter value is closer to that found towards the Sgr\,B2\,(M) hot core \citep{Nummelin00}. This suggests that this molecule might arise from the very dense hot cores in starburst regions. However, our data can neither support, nor refute this possibility.

\item{\bf CH$_2$NH}. Methyleneimine (or CH$_2$NH) is possibly formed by neutral-neutral gas-phase reactions, although grain-chemistry cannot be ruled out. This species is indeed poorly understood \citep{Turner99}. CH$_2$NH was observed to be more abundant in Galactic star-forming regions than in dark clouds \citep{Dickens97}. This indicates that this molecule is enhanced in warm environments, so it might be considered a good tracer of high temperatures. Outside the Milky Way, methyleneimine was detected towards NGC\,253 \citep{Martin06b} and the ULIRG galaxy Arp220 \citep{Salter08}. Our non-detection in M\,82 indicates that its abundance is at least a factor of 2.5 lower than in NGC\,253.

\item {\bf Deuterated molecules}. No deuterated molecules have been detected in M\,82. In accordance with the \citet{Bayet10} models, at least D$_2$CO, DCN, C$_2$D, and HDCS should be detectable in both M\,82 and NGC\,253. However, only DNC, and N$_2$D$^+$ are tentatively detected in NGC\,253 \citep{Martin06b}. Using the HNC data obtained by \citet{Huttemeister95} for M\,82 and our upper limits for DNC, we obtain ${\rm D/H}\le5\times10^{-4}$. Similarly, using the N$_2$H$^+$ data obtained by \citet{Sage95} we obtain ${\rm D/H}\le5\times10^{-2}$. The limit derived with DNC is consistent with the interstellar medium value of $(0.5-2)\times10^{-5}$ \citep{Turner78,Walmsley87,Mauersberger88,Parise09}.

\item{{\bf {Radio recombination lines}}} indicate the presence of ionized gas and can be used as an indicator of the star formation activity. There is a direct relationship between the integrated emission of the H\,II regions and the  OB star formation rate \citep{Cohen76,Kennicutt83}. Comparing our data and that of \citet{Martin06b}, we find that the integrated emission of the H$\alpha$ lines is more than 1.5 times higher in NGC\,253 than in M\,82. This indicates that there are more hot stars in the nucleus of NGC\,253. It could, however, also be related to the evolutionary stage of the star forming regions at the centres of both galaxies, where young stars emitting radiation are more common in a early-to-intermediate stage of bursts of star formation than in old starbursts.

\end{itemize}

\section{Conclusions: A chemical scenario of the M\,82 starburst}
\label{conclusions}
At first glance, Figs.~\ref{FigCompAbund} and ~\ref{FigCompAbundLimits} display striking differences between M\,82 and NGC\,253, with spectral lines (such as HNCO) varying by more than an order of magnitude. We have inferred that these differences are related to the heating mechanisms of the ISM in the central regions of both galaxies. Our observed deficiency of species such as CH$_3$OH, HNCO, NH$_3$, and SiO in M82 relative to other starburst galaxies has been reported previously \citep{Takano02,Martin06a,Martin09a}. In addition, we have found a deficiency of NS and HOCO$^+$. The ratio NS/CS (proposed as an indicator of shocks in Galactic hot cores, \citealt{Viti01}), is found to be an order of magnitude lower in M\,82 than in NGC 253. These abundance differences point towards a lack of dense molecular material in the central region of M82, together with enhanced heating and ionization due to UV radiation from newly formed OB stars. These enhanced UV fields leave a clear imprint in the molecular composition as illustrated by the overabundance of species such as CO$^+$, HCO, HOC$^+$, C$_3$H$_2$, and CH$_3$CCH  (reported by \citealt{Fuente05,Martin06b,Martin09b,Aladro10}). It is worth noting how the abundance of the PDR tracers CO$^+$, HCO, and HOC$^+$ relative to HCO$^+$ (which have been claimed to prove the significant presence of PDRs in NGC253, \citealt{Martin09b}) are similar in both M\,82 and NGC\,253. However, their relative abundances in dense molecular gas traced by C$^{34}$S show a significant overabundance in M\,82 as a result of the larger amount of gas affected by photodissociation. 

The high rotational temperatures measured with NH$_2$CN transitions in both M\,82 and NGC\,253 are similar to those derived in Galactic hot cores. If this molecule is tracing the very densest gas associated with the hot cores, a higher star-formation rate in M\,82 relative to that in NGC\,253 might be the cause of the observed slight enhancement of NH$_2$CN in the former. Although a significant amount of hot cores are present in M\,82, the dense molecular gas reservoir flowing into the central few hundred parsecs of this galaxy will be exhausted more quickly than in other starburst galaxies. However, since NH$_2$CN is so faint in M\,82, further observations of this molecule towards other extragalactic sources are needed to confirm its link with dense and hot regions.

Even if strong differences are found between M\,82 and NGC\,253, the comparative study based on the HNCO/CS ratio for a sample of galaxies by \citet{Martin09a} showed how NGC\,253 might be in a significantly evolved starburst stage, closer to the evolved stage of M\,82 than to the younger starbursts in galaxies such as M\,83. Thus, yet more varied molecular compositions are expected in a wider sample of galaxies including young starbursts.

\section{Summary}
\label{summary}

We have surveyed the NE molecular lobe of the M\,82 starburst galaxy between 129.8\,GHz and 175.0\,GHz and between 241.0\,GHz and 260.0\,GHz. In total, we have detected 69 spectral features corresponding to 18 different molecular species, as well as 3 hydrogen recombination lines. The molecules H$_2$S, CH$_3$CN, H$_2$CS, NO, and NH$_2$CN are detected for the first time in this galaxy (for the last two we have tentative detections). In addition, we have confirmed previous tentative detections of SO, SO$_2$, C$^{34}$S and H$^{13}$CN. 

Under the local thermodynamic equilibrium approximation, we have derived rotational temperatures and column densities of the various molecular species. Although a large number of molecules arise from dense and moderately warm gas ($\rm n\ge10^4\rm\, cm^{-3}$, $T_{\rm rot}=[10-30]$\,K), some of them seem to trace the cold gas contained in the outer envelopes of the molecular clouds (e.g. NO), while others might trace the very warm molecular gas phase (e.g. NH$_2$CN).

Making use of the frequency survey of \citet{Martin06b} carried out towards the galaxy NGC\,253, we have performed the most detailed comparison so far of the chemistry of two starburst galaxies, including twelve species that remain undetected in M\,82, but for which good upper limits could be obtained. We found that, although both galaxies have some similar characteristics that are typical of starburst galaxies, there are many species that point to different dominant processes. For example, C$_2$H, CH$_3$CCH, c-C$_3$H$_2$, or CO$^+$ are more abundant in  M\,82, while CH$_3$OH, HNCO, CH$_3$NH, NS, HOCO$^+$, and SiO are more abundant in NGC\,253.

We propose that the ratios $\rm NS/\rm C^{34}\rm S$ and $\rm HOCO^+/\rm C^{34}\rm S$ are new diagnostics for investigating the evolution of starbursts. These molecules, which are not detected in M\,82 but are  detected in NGC\,253, could provide valuable constrains on any ultraviolet fields or shocks in the ISM of starburst galaxies.

\begin{acknowledgements}
We thank the IRAM staff for their help with the observations R. Aladro acknowledges the hospitality of ESO Vitacura and the Joint ALMA Observatory. This work has been partially supported by the Spanish Ministerio de Ciencia e Innovaci\'on under projects ESP2007-65812-C02-01 and AYA2008-06181-C02-02.
\end{acknowledgements}

\clearpage

\begin{appendix}
\section{Comments on individual molecules}
\label{sect.molecules}
Eighteen molecular species were detected in our frequency survey between 129.8\,GHz and 175.0 \,GHz in the 2\,mm atmospheric window and between 241.0\,GHz and 260.0\,GHz in the 1.3\,mm window. In this appendix, we provide details of the Gaussian profile fitting and blending cases of individual transitions.

For some special cases, we performed synthetic Gaussian fits using the MASSA software, which reads the spectra and the spectroscopic parameters of the selected lines (using the JPL catalog), and then attemps to measure iteratively the column density, excitation temperature, velocity, and FWHM of the lines until it finds a simultaneous, optimal, Gaussian fits for all of them. 

For instance, the synthetic Gaussian fit method was applied to the weak molecules NO, NH$_2$CN, SO, and SO$_2$. For each species, all the transitions that lie in our survey were included, enabling MASSA to compare the strengths of the lines and ensure that the line identifications are reliable (since all the transitions of the same strength are either identified or not). 

In a similar way, synthetic Gaussian fits were applied to molecules with hyperfine structure. In these cases, MASSA fits one Gaussian profile to each component, taking into account the strength of each one. However, since the M82 lines are quite broad ($60-120$\,km\,s$^{-1}$), the hyperfine structure is not resolved and results in one single Gaussian (see \citealt{Martin10} for details about the C$_2$H$(2-1)$ synthetic fit). This method was also applied to these cases where two or more species are blended (see Figure~\ref{SyntheticGauss} for an example).

The synthetic Gaussian fits do not provide errors in the integrated intensities. However, these errors are required when doing the Boltzmann diagrams if one wishes to obtain appropriate errors associated with the rotational temperatures and column densities. Thus, unless otherwise indicated, in cases where a synthetic Gaussian profile was fitted, we calculated an error in the integrated area using 
\begin{equation}
(3\times rms \times FWHM)\,/\,(FWHM/Dv)^{1/2},
\end{equation}
where $rms$ is the 1$\sigma$ noise level of the spectrum that contains the line, and $Dv$ is the spectral resolution in velocity units. All the parameters are in international system units. 

\begin{itemize}
\item{\bf Hydrogen recombination lines}\\
The $\rm H36\alpha$, $\rm H35\alpha$, and  $\rm H34\alpha$ were observed at 135.3\,GHz, 147.0\,GHz, and 160.2\,GHz respectively. However, $\rm H35\alpha$ is blended with $\rm CS\,(3-2)$ and $\rm H36\alpha$ is blended with H$_2$CS\,$(4_{1,4}-3_{1,3})$. 
We first fitted a Gaussian profile to $\rm H34\alpha$, which is the only recombination line that does not suffer from contamination by other species. We then assumed that the three lines have similar intensities, and used the same Gaussian profile to subtract $\rm H35\alpha$ and $\rm H36\alpha$ from the blended spectra.

\item{\bf Cyanoacetylene - HC$_3$N}\\
We detected five transitions of this linear molecule. The HC$_3$N\,$(16-15)$ transition at 145.561\,GHz is blended with H$_2$CO\,$(2_{0,2}-1_{0,1})$ (Fig.~\ref{SyntheticGauss}). For this cyanoacetylene line, we fitted a synthetic Gaussian profile according to the physical parameters derived from all the other detected transitions, and fixed the velocity and the line width to 300\,km\,s$^{-1}$ and 100\,km\,s$^{-1}$, respectively. We also included the transition HC$_3$N\,$(24-23)$ at 218.325\,GHz observed by \citet{Aladro10}.

\item{\bf Propyne (methyl acetylene) - CH$_3$CCH}\\
None of the five detected lines of this symmetric top molecule are blended. Each  $(J\rightarrow J-1)$ transition consists of a number of $K$ components (being $K=0,\dots,J-1$) that are not resolved because of the large line widths. Thus, we fitted unique Gaussian profile to each group of transitions. Given the rotational temperature derived from the Boltzmann diagram ($T_{\rm rot}=31.6\pm0.1$\,K), only the $K=0$ to 4 components contribute to the line intensity, with the  $K=4$ component having a contribution of less than 1\% to the total $T_{\rm peak}$. Therefore, higher values of the $K$-ladder (i.e., $K>4$) were not  taken into account. 

Before fitting the Gaussian, we subtracted a bump in the CH$_3$CCH\,$(8-7)$ line due to emission from the centre of M\,82, that was detected by the beam of the telescope. The CH$_3$CCH\,$(9-8)$ transition, at 153.817\,GHz could be contaminated by the HNCO\,$(7_{0, 7}-6_{0,6})$ line, but because no other lines of isocyanic acid with similar spectroscopic parameters are detected, we assume that its contribution is  negligible. From \citet{Aladro10} we gathered data for the CH$_3$CCH\,$(16-15)$ transition at 273.420\,GHz, which lies outside our survey coverage. 

\item{\bf Carbon monosulfide - CS}\\
Two transitions of CS fall within our frequency survey. The line CS\,$(3-2)$ at 146.969\,GHz is blended with $\rm H35\alpha$. We first subtracted the estimated contribution of the hydrogen recombination line, and then fitted a Gaussian profile to the residual feature. The CS\,$(5-4)$ at 244.936\,GHz does not show any special characteristics. To estimate the physical parameters of CS, we also used the detections reported by \citet{Bayet09b} and \citet{Aladro10}. This is the only molecule in the survey that haves two components in the Boltzmann diagram, with rotational temperatures of $5.8\pm0.1$\,K and $15.1\pm0.9$\,K \citep{Aladro10}.

\item{\bf Methyl cyanide - CH$_3$CN}\\
Three transitions of methyl cyanide were clearly detected at 147.174\,GHz, 165.565\,GHz, and 257.483\,GHz, none of which are blended. In a similar way to propyne, CH$_3$CN is a symmetric top molecule whose lines contain a number of unresolved $K$ components $(K=0,\dots,J-1)$. As a first approximation, the Boltzmann diagram gave us $T_{\rm rot}\sim$\,30\,K. For this temperature only the $K=0,\dots,4$ components contribute to the total line intensities, the $K>4$ components contributing less than 1\%. Taking into account only these first $K-$ladder components, we finally obtain a $T_{\rm rot}=32.9\pm1.6$\,K. 

On the other hand, as mentioned in Sect.~\ref{sect.comp} this molecule probably arises from the cores of the molecular clouds. This is consistent with  the narrow FWHM of its lines, about 80\,kms$^{-1}$ for CH$_3$CN\,$(8_K-7_K)$ and CH$_3$CN\,$(9_K-8_K)$, and especially for the $\sim$60\,kms$^{-1}$ of the CH$_3$CN\,$(14_K-13_K)$ transition.

\item{\bf Hydrogen cyanide - H$^{13}$\,CN}\\
The group of transitions of H$^{13}$CN\,$(2_K-1_K)$ were detected at 172.678\,GHz. We fitted a synthetic Gaussian profile using the same line width and position for the five hyperfine components of the line. We then assumed $T_{\rm rot}=20\pm10\,$K in the Boltzmann diagram. The large error in the column density reflects the error in the integrated areas found using the formula in Eq. A1.

Following the lower limit to the carbon isotopic ratio in M\,82, $^{12}\rm C/^{13}\rm C>138$ by \citet{Martin10}, we can estimate the column density of HCN as $N_{\rm HCN}>2.3\times10^{14}$\,cm$^{-2}$, in agreement with the results of \citet{Seaquist00}.

\item{\bf Formaldehyde - H$_2$CO}\\
We detected three transitions of this molecule, the H$_2$CO\,$(2_{1,2}-1_{1,1})$ line at 140.840\,GHz being the only that is not blended. Thus, the width and position of the Gaussian fit of this line were used to fix the Gaussian fitting to the other two lines. The transition H$_2$CO\,$(2_{0,2}-1_{0,1})$ at 145.603\,GHz is blended with HC$_3$N\,$(16-15)$, and is also contaminated by the emission from the centre of the galaxy detected by the telescope beam (Fig.~\ref{SyntheticGauss}). We first subtracted the contributions of both cyanoacetylene and the emission from the centre of the galaxy, and then fitted a Gaussian profile to the residuals. Likewise, H$_2$CO\,$(2_{1,1}-1_{1,1})$ at 150.498\,GHz is blended with c-C$_3$H$_2\,(2_{2,0}-1_{1,1})$ and NO\,$(3/2-1/2)\Pi^-$. In this case, we first fitted a Gaussian profile to the formaldehyde feature. All the fits are synthetic.

\begin{figure}
\begin{center}
\includegraphics[width=0.5\textwidth]{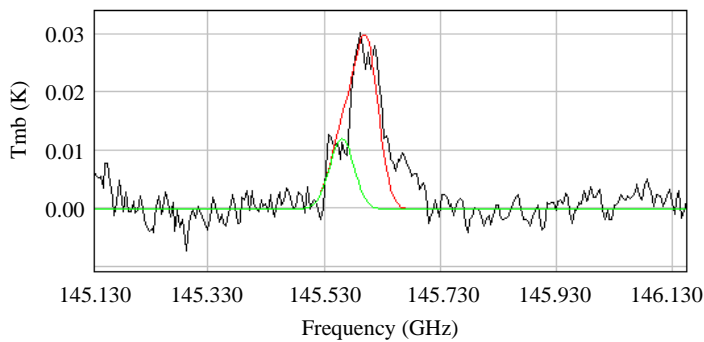}
\caption{Example of a synthetic Gaussian fit performed with MASSA. In this case, HC$_3$N\,$(16-15)$ (the line on the left) is blended with H$_2$CO\,$(2_{0,2}-1_{0,1})$ (the line to the right). The bump at higher frequencies is the emission from the centre of the galaxy picked up by the telescope beam. This bump was originally subtracted before any Gaussian fit, although here we show it as an example of the strength of this emission. See details about the Gaussian fitting in the text.}
\label{SyntheticGauss}
\end{center}
\end{figure}

\item{\bf Cyclopropenylidene - c-C$_3$H$_2$}\\
A total of eight transitions of this molecule were detected in the survey. Six of them are blended with other lines, and two are tentative. The transitions c-C$_3$H$_2\,(3_{1,2}-2_{2,2})$ at 145.090\,GHz and c-C$_3$H$_2\,(2_{2,0}-1_{1,1})$ at 150.436\,GHz are blended with CH$_3$OH\,$(3-2)$A+ and H$_2$CO\,$(2_{1,1}-1_{1,1})$, respectively. Both Gaussian profile fits are synthetic. The pair of lines at 150.8\,GHz is spectrally unresolved. The same holds for the pair at 151.3\,GHz. In all these cases, the width and the velocity of the cyclopropenylidene Gaussian profiles were fixed to 100.0 km\,s$^{-1}$ and 300 km\,s$^{-1}$, respectively. The fixed velocity is supported by the results of the two non-blended lines, at 305.1 and 307.6 km\,s$^{-1}$. However, only the c-C$_3$H$_2\,(3_{2,1}-2_{1,2})$ line at 244.222\,GHz has a non-fixed line width, which is a bit narrower than 100\,km\,s$^{-1}$ ($\sim$85\,km\,s$^{-1}$). Some molecules occasionally have narrower features at higher frequency transitions (e.g. CS, \citealt{Bayet09b}), hence the assumption of a line width of 100\,km\,s$^{-1}$ may be a good guess for the lower transitions. The tentative detections are the two at 151.3\,GHz, with line intensities of $\sim$1 and $\sim$2\,mK. 

\item{\bf Methanol - CH$_3$OH}\\
Six lines or groups of lines of methanol were identified at 145.103, 157.179, 157.271, 165.050, 170.060, and 241.791\,GHz. A synthetic Gaussian profile was fitted to all of them since, except for the lines at 170.060\,GHz and 241.791\,GHz, all the transitions are tentative, having an intensity lower than two times the noise level of the corresponding spectrum. In addition, the CH$_3$OH\,$(3-2)$A+ line at 145.103\,GHz is blended with c-C$_3$H$_2\,(3_{1,2}-2_{2,2})$, whose contribution had been previously subtracted. The group of  CH$_3$OH\,$(J_{1,K-1}-J_{0,K})$E lines at 165.050\,GHz is also blended with the faint feature of SO$_2\,(5_{2,4}-5_{1,5})$. In this case, we first estimated the contribution of the methanol lines using their spectroscopic parameters and helping us with the other non-blended transitions.

From the Boltzmann diagram, we obtain $N_{\rm CH_3OH}=(1.2\pm1.1)\times10^{14}$\,cm$^{-2}$ and $T_{\rm rot}=26.2\pm11.7$\,K. The large errors are related to the low intensity of the lines because the baselines of the spectra have a strong influence on the results of the rotational parameters.

\item{\bf Nitric oxide - NO}\\
Four transitions of nitric oxide were tentatively detected at 150.176, 150,546, 250.437, and 250.796\,GHz. Although they are clearly seen (in particular the two at 2\,mm wavelengths), their intensities do not reach a S/N of 2. The line at 150.546\,GHz is blended with H$_2$CO\,$(2_{1,1}-1_{1,1})$, whose contribution was firstly subtracted as explained for formaldehyde. We fitted synthetic Gaussian profiles to the 2\,mm and 1.3\,mm hyperfine lines separately. Since we had two groups of lines at different frequencies, we were able to calculate the rotational temperature. However, this value is not very precise because the dynamic range in energies is quite low ($\Delta \rm E=5\,\rm cm^{-1}$). Thus, we associated a 50\% error to the derived rotational temperature. The low $T_{\rm rot}\sim9$\,K  and the high column density ($N\sim10^{15}\rm cm^{-2}$) indicate that nitric oxide arises from the external cold layers of the molecular clouds.

\item{\bf Thioformaldehyde - H$_2$CS}\\
Only one transition of thioformaldehyde was detected at 135.297\,GHz, which is blended with H$36\alpha$. We first subtracted the contribution of the recombination line and then fitted a Gaussian profile to the residuals. Since we only detected one transition of this species in the whole survey, we fixed the rotational temperature to $20\pm10$\,K to calculate its column density.%

\item{\bf Ethynyl radical - C$_2$H}\\
Only the C$_2$H\,$(2-1)$ transition falls within our frequency coverage. It is the brightest spectral feature and the most abundant molecule of the survey. This line also shows the contribution of emission from the centre of M\,82, which was firstly subtracted. After that, we fitted a synthetic Gaussian profile to the unresolved hyperfine structure, taking into account the spectroscopic parameters and line intensities of each component. Since we had only one transition of this molecule, we assumed $T_{\rm rot}=20\pm10$\,K. A study dedicated to this line in M\,82 and NGC\,253 is presented in \citet{Martin10}.

\item{\bf  Sulphur monoxide - SO}\\
Only the SO\,$(4_3-3_2)$ transition was detected in the survey at 138.179\,GHz. Fixing $T_{\rm rot}$  to $20\pm10$\,K, we obtained a column density of $N_{\rm SO}=(3.7\pm0.5)\times10^{13}$\,cm$^{-2}$.

\item{\bf Sulphur dioxide - SO$_2$}\\
Only the SO$_2\,(5_{2,4}-5_{1,5})$ transition at 165.145\,GHz was detected in the survey. It is blended with a methanol group of lines whose contribution was firstly subtracted as explained before. After fitting a synthetic Gaussian profile to the residuals with a fixed position of 300\,km\,s$^{-1}$, the SO$_2$ line intensity was estimated to be $\sim$1\,mK. This value is below the noise level at this frequency so the detection is considered as tentative. Because this line is so faint, the Gaussian parameters are strongly affected by the baseline. As mentioned at the beginning of this section, the synthetic fit does not provide any error in the integrated area of the line. In this case, we did not use equation A1 to calculate the error in the integrated area, since the error is even higher than $\int T_{\rm MB}$\,dv. Thus, to calculate the column density we assumed a rotational temperature of $20\pm10$\,K, and assigned an error of 50\% to the resulting $N_{\rm SO_2}$.

\item{\bf Oxomethyl - HCO}\\
We detected the HCO\,$(2_{0,2}-1_{0,1})$ transition at 173.377\,GHz, which contains seven hyperfine lines with lower level energies of 2.9\,K. It is blended with H$^{13}$CO\,$^+(2-1)$. We first fitted a synthetic Gaussian profile to the hyperfine lines at 173.377 and 173.406\,GHz because they are less blended with H$^{13}$CO$^+$. We fixed the position and line width to 300\,km\,s$^{-1}$ and 100\,km\,s$^{-1}$ respectively. In this way, we made sure that the synthetic fit covered oxomethyl and not H$^{13}$CO$^+$. We then fitted the rest of the hyperfine lines with the same Gaussian parameters. To calculate the column density, we used a rotational temperature of $20\pm10$\,K. As in the case of SO$_2$, the error in the integrated area from equation A1 is larger than the area itself, thus we did not use any error in the Boltzmann diagram. We note, however, that the resulting error in the column density is quite large.

\item{\bf Oxomethylium, formyl cation - H$^{13}$CO$^+$}\\
The H$^{13}$CO$^+\,(2-1)$ line at 173.507\,GHz is blended with oxomethyl. After fitting a Gaussian profile to the HCO feature, we over-fitted another to the residuals, fixing the position and line width to 300\,km\,s$^{-1}$ and 100\,km\,s$^{-1}$. Using the carbon isotopic ratio in M\,82 given by \citet{Martin10}, $^{12}\rm C/^{13}\rm C>138$, we estimated a column density of HCO$^+$ of $N_{\rm HCO^+} > 1.8\times10^{14}$\,cm$^{-2}$. This value is consistent with that derived by \citet{Seaquist00}.

\item{\bf Cyanamide - NH$_2$CN}\\
Five spectral features of cyanamide were tentatively detected between 158.815 and 158.943\,GHz. The line widths and positions in MASSA were fixed to 100.0 and 302.1\,km\,s$^{-1}$, respectively, which are the values previously obtained with an approximated non-synthetic Gaussian fit. The line intensities, as calculated with the synthetic Gaussian fitting, are about 2\,mK. The resulting Boltzmann diagram gave us $T_{\rm rot}=80.3\pm52.9$\,K and $N_{\rm NH_2CN}=1.2\pm1.5\times10^{13}$\,cm$^{-2}$. The errors in both parameters, in particular in the column density, reflect the strong effect of the baseline on this weakly emitting species. 

\item{\bf Hydrogen sulfide - H$_2$S}\\
The H$_2$S\,$(1_{1,0}-1_{0,1})$ transition was detected at 168.763\,GHz. Since there is only one transition of this species, we used a $T_{\rm rot}=20\pm10$\,K in the rotational diagram and derived a column density of $N_{\rm H_2S}=(6.1\pm4.5)\times10^{13}$cm$^{-2}$.

\end{itemize}

\clearpage

\Online

\onecolumn
\begin{figure*}[!ht]
\begin{center}

\includegraphics[width=\textwidth]{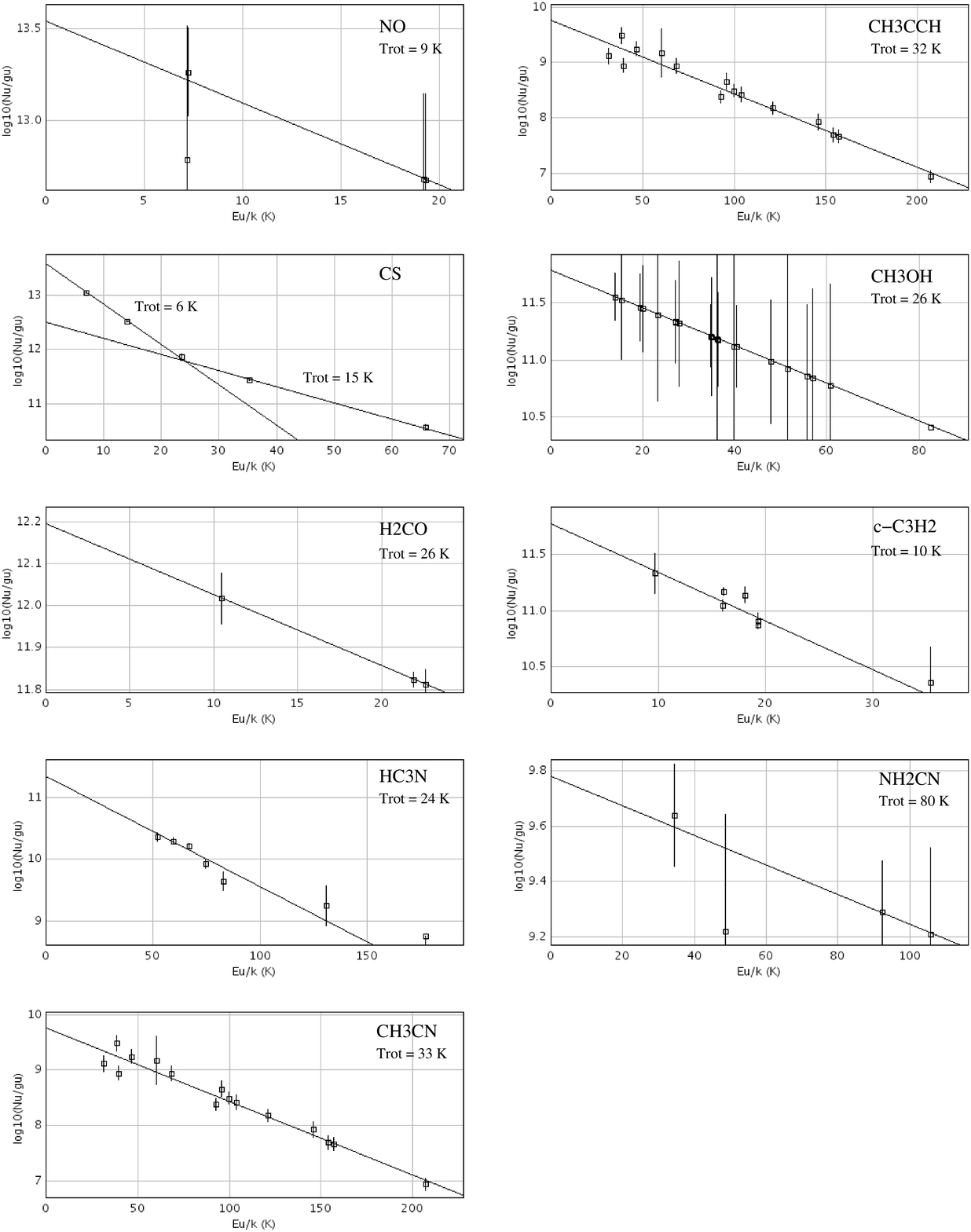}

\caption{Boltzmann diagrams of the molecules observed in the survey for which we detected more than one transition. The resulting rotational temperatures are indicated.}
\label{FigDR}
\end{center}
\end{figure*}

\clearpage

\onecolumn

\begin{longtable}{l c c c c c c c c c }

\hline
\hline
 Line	&  Frequency   & $\int$ $T_{\rm{MB}}$\,dv	& $V_{\rm{LSR}}$ &  FWHM	& $T_{\rm{MB}}$  &  Remarks\\
    & GHz & K\,km\,s$^{-1}$     & km\,s$^{-1}$ & km\,s$^{-1}$   &  K  &  \\[0.2cm]
\hline
H$36\,\alpha$ 			& 135.286 &  1.07\,$\pm$\,0.00 & 325.0\,$\pm$\,0.0  & 80.0\,$\pm$\,0.0& 0.012 & b \\
H$_2$CS\,$(4_{1,4}-3_{1,3})$ 	& 135.297 & 0.51\,$\pm$\,0.09   & 291.6\,$\pm$\,7.8 & 68.7\,$\pm$\,10.5 & 0.006 & b \\
HC$_3$N\,$(15-14)$ 		& 136.464 & 1.19\,$\pm$\,0.18 & 308.8\,$\pm$\,8.0 & 97.3\,$\pm$\,17.0 & 0.012 & \\
CH$_3$CCH\,$(8_K-7_K)$ 		& 136.728 & 7.4\,$\pm$\,0.6  &  310.0\,$\pm$\,0.0 & 122.9\,$\pm$\,12.1 & 0.057 & m \\
SO\,$(4_3-3_2)$ 		& 138.179 & 0.94\,$\pm$\,0.13 & 313.3\,$\pm$\,6.5 &  93.2\,$\pm$\,15.2 &  0.009 &\\ 
H$_2$CO\,$(2_{1,2}-1_{1,1})$ 	& 140.840 & 4.36\,$\pm$\,... & 300.3\,$\pm$0.0 &   101.1\,$\pm$\,0.0 & 0.040 & s  \\ 
C$^{34}$S\,$(3-2)$ 		& 144.617 & 1.0\,$\pm$\,0.5 & 291.2\,$\pm$\,28.8 &  122.3\,$\pm$\,79.6 & 0.008 & \\ 
c-C$_3$H$_2\,(3_{1,2}-2_{2,2})$ & 145.090 & 1.48\,$\pm$\,... & 300.0\,$\pm$\,0.0 &  100.0\,$\pm$\,0.0 & 0.014  & b, s\\ 
CH$_3$OH\,$(3-2)$ A+ 		& 145.103 & 0.26\,$\pm$\,... & 330.0\,$\pm$\,0.0 & 80.0\,$\pm$\,0.0 & 0.003 & m, b, s\\	
HC$_3$N\,$(16-15)$ 		& 145.561 & 1.29\,$\pm$\,... & 300.0\,$\pm$\,0.0 &  100.0\,$\pm$\,0.0 & 0.012 & b, s \\ 
H$_2$CO\,$(2_{0,2}-1_{0,1})$ 	& 145.603 & 3.31\,$\pm$\,... & 300.3\,$\pm$\,0.0 &  101.1\,$\pm$\,0.0 & 0.031 & b, s\\
CS\,$(3-2)$ 			& 146.969 & 11.48\,$\pm$\,0.19 &  293.8\,$\pm$\,0.9 &  105.6\,$\pm$\,2.1 &    0.102 & b\\ 	
H\,$35\,\alpha$ 		& 147.047 & 1.07\,$\pm$\,0.0 &  325.0.0\,$\pm$\,0.0 &   80.0\,$\pm$\,0.0 & 0.012 & b\\ 
CH$_3$CN\,$(8_K-7_K)$ 		& 147.174 & 0.55\,$\pm$\,0.17 & 300.0\,$\pm$\,0.0 &  80.0\,$\pm$\,0.0 &  0.006 & m\\
NO\,$(3/2-1/2)\Pi$+ 		& 150.176 & 0.10\,$\pm$\,...     & 307.7\,$\pm$\,0.0 & 114.2\,$\pm$\,0.0  &  0.002 & hf, s\\
c-C$_3$H$_2$\,$(2_{2,0}-1_{1,1})$ & 150.436 &  0.57\,$\pm$\,... &  300.0\,$\pm$\,0.0 & 100.0\,$\pm$\,0.0 &   0.005 & b, s\\
H$_2$CO\,$(2_{1,1}-1_{1,1})$ 	& 150.498 &   4.98\,$\pm$\,... & 300.3\,$\pm$\,0.0 & 101.1\,$\pm$\,0.0 &   0.046 & b, s \\
NO\,$(3/2-1/2)\,\Pi$- 		& 150.546 &   0.29\,$\pm$\,... &  307.7 \,$\pm$\,0.0 &  114.2\,$\pm$\,0.0 &  0.002 & hf, b, s \\
c-C$_3$H$_2$\,$(4_{0,4}-3_{1,3})$ & 150.821 &  1.04\,$\pm$\,0.16 &  300.0\,$\pm$\,0.0 & 100.0\,$\pm$\,0.0 & 0.010 & b \\
c-C$_3$H$_2$\,$(4_{1,4}-3_{1,3})$ & 150.851 &  2.91\,$\pm$\,0.17 & 300.0\,$\pm$\,0.0 &  100.0\,$\pm$\,0.0 & 0.027 & b\\
c-C$_3$H$_2$\,$(5_{1,4}-5_{0,5})$ & 151.344 & 0.26\,$\pm$\,0.06 & 300.0\,$\pm$\,0.0 &  100.0\,$\pm$\,0.0 & 0.002 & b\\
c-C$_3$H$_2$\,$(5_{2,4}-5_{1,5})$ & 151.361 & 0.09\,$\pm$\,0.06 & 300.0\,$\pm$\,0.0 & 100.0\,$\pm$\,0.0 & 0.001 & b\\
CH$_3$CCH\,$(9_K-8_K)$ 		& 153.817 &  7.46\,$\pm$\,0.16 &  322.6\,$\pm$\,1.0 &  109.6\,$\pm$\,2.8 &  0.060 & m \\
HC$_3$N\,$(17-16)$ 		& 154.657 &  1.30\,$\pm$\,0.14 &  314.1\,$\pm$\,5.7 &  95.4\,$\pm$\,10.2 & 0.013 &\\
c-C$_3$H$_2$\,$(3_{2,2}-2_{1,1})$ & 155.518 &  0.98\,$\pm$\,0.08	& 305.1\,$\pm$\,4.3 &  100.0\,$\pm$\,0.0 &   0.009 & \\	
CH$_3$OH\,$(5_{0,5}-5_{-1,5})$\,E & 157.179 & 0.08\,$\pm$\,... & 326.6\,$\pm$\,0.0    &  63.2\,$\pm$\,0.0    & 0.001 & s\\
CH$_3$OH\,$(J_{0,K}-J_{-1,K})$\,E & 157.271 & 0.12\,$\pm$\,... & 326.6\,$\pm$\,0.0    &  63.2\,$\pm$\,0.0     & 0.002 & m, s\\
NH$_2$CN\,$(8_{1,7}-7_{0,7})$ 	& 159.815 &   0.26\,$\pm...$ & $302.1\,\pm0.0$   &  100.0\,$\pm$\,0.0 &  0.002 & s\\
NH$_2$CN\,$(8_{1,8}-7_{1,7})$	& 158.891 & 0.26$\pm...$      & $302.1\pm0.0$        &  100.0\,$\pm$\,0.0 & 0.002  & s \\
NH$_2$CN\,$(8_{1,7}-7_{0,7})$ 	& 159.943 &  0.24\,$\pm...$ & $302.1\,\pm0.0$   & 100.0\,$\pm$\,0.0   & 0.002 & m, s\\
H$34\,\alpha$  			& 160.211 &  1.07\,$\pm$\,0.12 & 325.0\,$\pm$\,5.8    &  80.0\,$\pm$\,0.0 & 0.012 &\\
HC$_3$N\,$(18-17)$ 		& 163.753 & 0.80\,$\pm$\,0.13 & 308.5\,$\pm$\,7.5    &  87.5\,$\pm$\,14.5&  0.009 &\\
CH$_3$OH\,$(J_{1,K-1}-J_{0,K})$\,E & 165.050 & 0.13\,$\pm$\,..  & 330.0\,$\pm$\,0.0  &  80.0\,$\pm$\,0.0 & 0.002 & m, b, s\\
SO$_2$\,$(5_{2,4}-5_{1,5})$   	& 165.145 & 0.09\,$\pm$\,... & 300.0\,$\pm$\,0.0    &  78.3\,$\pm$\,0.0   &  0.001 & b, s\\
CH$_3$CN\,$(9_K-8_K)$  		& 165.569 &  0.48\,$\pm$\,0.14 & 268.4\,$\pm$\,12.2   &  85.7\,$\pm$\,31.2 &  0.005 & m \\
H$_2$S\,$(1_{1,0}-1_{0,1})$    	& 168.763 &  2.67\,$\pm$\,0.18   &  310.4\,$\pm$\,2.7 & 80.6\,$\pm$\,6.6  &  0.031 &  \\  
CH$_3$OH\,$(3_{2,1}-2_{1,1})$ 	& 170.060 &  0.48\,$\pm$\,... &  260.1\,$\pm$\,0.0     & 61.3 \,$\pm$\,0.0 &  0.007 & s  \\  
CH$_3$CCH\,$(10_K-9_K)$ 	& 170.906 & 9.06\,$\pm$\,0.18 & 324.9\,$\pm$\,1.0    & 100.0\,$\pm$\,2.4 &  0.085 & m\\
H$^{13}$CN\,$(2_K-1_K)$ 	& 172.678 &  0.67\,$\pm$\,0.15 & 319.8\,$\pm$\,10.4   &  90.3\,$\pm$\,21.3 &  0.007 & hf, s\\ 
HC$_3$N\,$(19-18)$  		& 172.849 &  0.50\,$\pm$\,0.18 & 324.1\,$\pm$\,18.8   &  95.8\,$\pm$\,41.0 & 0.005 & \\
HCO\,$(2_{0,2}-1_{0,1})$    	& 173.377 & 0.86\,$\pm$\,...  & 300.0\,$\pm$\,0.0    & 100.0\,$\pm$\,0.0  & 0.008 & b, hf, s\\
H$^{13}$CO$^+$\,$(2-1)$		& 173.507 & 0.90\,$\pm$\,... & 300.0\,$\pm$\,0.0 & 100.0\,$\pm$\,0.0 & 0.008 & b, s \\	
C$_2$H\,$(2-1)$   		& 174.663 & 12.11\,$\pm$\,... & 307.5\,$\pm$\,0.0   &  95.6\,$\pm$\,0.0 & 0.119 & hf, s\\
CH$_3$OH\,$(5_{0,5}-4_{0,4})$\,A+  & 241.791 & 0.64\,$\pm$\,... & 343.7\,$\pm$\,0.0    & 93.4\,$\pm$\,0.0  & 0.006 & m, s \\
c-C$_3$H$_2$\,$(3_{2,1}-2_{1,2})$  & 244.222 & 0.96\,$\pm$\,0.16 & 307.6\,$\pm$\,7.3    & 84.5\,$\pm$\,12.6 & 0.011 & \\ 
CS\,$(5-4)$           		& 244.936 &  4.79\,$\pm$\,0.19 & 316.1\,$\pm$\,1.6    & 77.1\,$\pm$\,3.6  & 0.058 & \\
NO\,$(5/2-3/2)\Pi+$       		& 250.437 &  0.34\,$\pm$\,...      & 325.4\,$\pm$\,0.0    & 85.1\,$\pm$\,0.0  & 0.004 & hf, s\\
NO\,$(5/2-3/2)\Pi-$       		& 250.796 &  0.34\,$\pm$\,...      & 325.4\,$\pm$\,0.0    & 85.1\,$\pm$\,0.0  & 0.004 & hf, s\\
CH$_3$CCH\,$(15_K-14_K)$  	& 256.317 &  7.18\,$\pm$\,0.13 & 312.5\,$\pm$\,0.8    & 94.6\,$\pm$\,2.1  & 0.071 & m \\
CH$_3$CN\,$(14_K-13_K)$   	& 257.483 &  0.35\,$\pm$\,0.09 & 297.3\,$\pm$\,7.8    & 60.3\,$\pm$\,18.0 & 0.005  & m \\
\hline
H$_2$CO$(2_{1,2}-1_{1,1})$	& 140.840 &  $0.12\pm0.09$ &$300.0\pm0.0$  &  $100.0\pm0.0$ & 0.001 &  i \\
HCO$^+(3-2)$	& 267.557 & $4.18\pm0.27$ & $278.7\pm3.2$ & $102.4\pm7.8$ & 0.038 & i \\

\hline
\caption{Gaussian parameters for the fit performed to the observed spectral 
features. In those cases where the associated errors are zero, the parameter was fixed. Synthetic fits do not give errors in the integrated area. Remarks: $(b)$ blended line; $(m)$ multi-transition line. The parameters refer to the main component of the group; $(s)$ synthetic Gaussian fit using MASSA; $(hf)$ hyperfine transition; $(i)$ line coming from the image band. See Appendix A for more details.}
\label{TableGausParameters}
\end{longtable}
\end{appendix}

\end{document}